\documentclass[fleqn,usenatbib]{mnras}

\usepackage{newtxtext,newtxmath}
\usepackage[T1]{fontenc}
\usepackage{ae,aecompl}

\usepackage{graphicx}	% Including figure files
\usepackage{amsmath}	% Advanced maths commands
\usepackage{float}
\usepackage{subcaption}
\usepackage{geometry}
\usepackage{xcolor}
\newcommand{\Ms}{{\ensuremath{{M}_{\odot} }}}

\newcommand{\tff}{{\ensuremath{t_\mathrm{ff}}}}
\newcommand{\nH}{{\ensuremath{n_\mathrm{H}}}}
\newcommand{\nHcrit}{{\ensuremath{n_\mathrm{H, crit}}}}
\newcommand{\meannH}{{\ensuremath{\bar{n}_\mathrm{H}}}}

\newcommand{\meanNH}{{\ensuremath{\bar{N}_\mathrm{H}}}}

\newcommand{\meanNNNHp}{{\ensuremath{\bar{N}_\mathrm{N_2H^+}}}}

\newcommand{\C}{{\ensuremath{\mathrm{C}}}}
\newcommand{\CO}{{\ensuremath{\mathrm{CO}}}}

\newcommand{\Fe}{{\ensuremath{\mathrm{Fe}}}}

\newcommand{\He}{{\ensuremath{\mathrm{He}}}}
\newcommand{\Hy}{{\ensuremath{\mathrm{H}}}}
\newcommand{\HH}{{\ensuremath{\mathrm{H_2}}}}
\newcommand{\N}{{\ensuremath{\mathrm{N}}}}
\newcommand{\Ox}{{\ensuremath{\mathrm{O}}}}
\newcommand{\Na}{{\ensuremath{\mathrm{Na}}}}
\newcommand{\Mg}{{\ensuremath{\mathrm{Mg}}}}
\newcommand{\Cl}{{\ensuremath{\mathrm{Cl}}}}
\newcommand{\Ph}{{\ensuremath{\mathrm{Ph}}}}
\newcommand{\F}{{\ensuremath{\mathrm{F}}}}

\newcommand{\So}{{\ensuremath{\mathrm{So}}}}
\newcommand{\Si}{{\ensuremath{\mathrm{Si}}}}
\newcommand{\NNHp}{{\ensuremath{\mathrm{N_2H^+}}}}
\newcommand{\NNDp}{{\ensuremath{\mathrm{N_2D^+}}}}

\newcommand{\DfracNNH}{{\ensuremath{\mathrm{D^{\NNHp}_{\text{frac}}}}}}
\newcommand{\OPRHH}{{\ensuremath{\mathrm{OPR^{H_2}}}}}

\newcommand{\KROME}{\texttt{KROME}}
\newcommand{\ENZO}{\texttt{ENZO}}
\newcommand{\GANDALF}{\texttt{GANDALF}}

\title[Deuterium Chemodynamics of Massive PSCs]{Deuterium Chemodynamics of Massive Pre-Stellar Cores}

\author[C. J. Hsu et al.]{Chia-Jung Hsu,$^{1}$
Jonathan C. Tan,$^{1,5}$
Matthew D. Goodson,$^{2}$
Paola Caselli,$^{3}$
\newauthor
Bastian K\"{o}rtgen,$^{4}$
and Yu Cheng$^{5}$
\\
$^{1}$Department of Space, Earth \& Environment, Chalmers University of Technology, Gothenburg, Sweden \\
$^{2}$Dept. of Physics and Astronomy, University of North Carolina at Chapel Hill, Chapel Hill, NC 27599-3255, USA \\
$^{3}$Max-Planck-Institute for Extraterrestrial Physics (MPE), Giessenbachstr. 1, D-85748 Garching, Germany \\
$^{4}$Hamburger Sternwarte, Universit\"{a}t Hamburg, Gojenbergsweg 112, D-21029 Hamburg, Germany \\
$^{5}$Dept. of Astronomy, University of Virginia, Charlottesville, Virginia 22904, USA
}

\pubyear{2020}

\begin{document}
\label{firstpage}
\pagerange{\pageref{firstpage}--\pageref{lastpage}}
\maketitle

\begin{abstract}
High levels of deuterium fractionation of {\NNHp} (i.e., {\DfracNNH}$\gtrsim 0.1$) are often observed in pre-stellar cores (PSCs) and detection of {\NNDp} is a promising method to identify elusive massive PSCs. However, the physical and chemical conditions required to reach such high levels of deuteration are still uncertain, as is the diagnostic utility of {\NNHp} and {\NNDp} observations of PSCs. We perform 3D magnetohydrodynamics simulations of a massive, turbulent, magnetised PSC, coupled with a sophisticated deuteration astrochemical network. Although the core has some magnetic/turbulent support, it collapses under gravity in about one freefall time, 
%with central regions reaching high densities expected to immediately precede protostar formation in about 60~kyr, 
which marks the end of the simulations. Our fiducial model achieves relatively low {\DfracNNH}$\sim0.002$ during this time. We then investigate effects of initial ortho-para ratio of {\HH} ({\OPRHH}), temperature, cosmic-ray (CR) ionization rate, {\CO} and N-species depletion factors and prior PSC chemical evolution. We find that high CR ionization rates and high depletion factors allow the simulated {\DfracNNH} and absolute abundances to match observational values within one freefall time. For {\OPRHH}, while a lower initial value helps the growth of {\DfracNNH}, the spatial structure of deuteration is too widespread compared to observed systems. For an example model with elevated CR ionization rates and significant heavy element depletion, we then study the kinematic and dynamic properties of the core as traced by its {\NNDp} emission. The core, undergoing quite rapid collapse, exhibits disturbed kinematics in its average velocity map. Still, because of magnetic support, the core often appears kinematically sub-virial based on its {\NNDp} velocity dispersion. 
\end{abstract}


%%%%%%%%%%%%%%%%%%%%%%%%%%%%%%%%%%%%%%%%%%%%%%%%%%

%%%%%%%%%%%%%%%%% BODY OF PAPER %%%%%%%%%%%%%%%%%%

\section{Introduction}\label{S:introduction}

Massive stars play an important role in the universe. Their radiative, mechanical and chemical feedbacks influence the evolution of interstellar media and thus star formation and galactic evolution. However, the formation mechanism of massive stars is still unclear. Their short lifetimes and relative rarity make observational studies challenging. Candidate theories to explain massive star formation include Turbulent Core Accretion and Competitive Accretion. Turbulent Core Accretion \citep[e.g.,][]{2003ApJ...585..850M} proposes that a combination of supersonic turbulence and magnetic fields supports massive pre-stellar cores (PSCs) against fragmentation and that these then collapse to a central star-disk system. However, the collapse is not necessarily as ordered as in the case of low-mass star formation, especially if there is significant turbulence in the PSC. The Competitive Accretion model of \citet{2001MNRAS.323..785B} proposes that large numbers of low-mass stars form in a protocluster clump, with a few of them later accreting chaotically, by Bondi-Hoyle accretion of gas supplied by the collapsing clump, to become massive stars. To distinguish these two theories, one of the decisive differences is in the existence of massive, coherent PSCs.

One way to identify PSCs and also characterise their dynamical history is via the deuterium fractionation of certain chemical tracers. In particular, {\NNHp} and its deuterated form, {\NNDp}, have been observed in PSCs \citep[e.g.,][]{Caselli2002,Tan2013,Kong2016}. Due to the cold ($T<20\:$K) and dense ($\mathrm{n_H} > 10^5\:{\rm cm}^{-3}$) environments of PSCs, the deuteration process gradually increases the deuterium fraction of {\NNHp} (\DfracNNH $\equiv [\NNDp]/[\NNHp]$). Also, as deuteration of {\NNHp} is sensitive to the ortho-to-para ratio (OPR) of {\HH}, which is expected to gradually decline after formation of molecular gas, \DfracNNH could be a proxy to estimate the chemical age of PSCs.

% To probe the ages of pre-stellar cores, the measurement of the molecular deuteration is believed as an effective method. Molecular deuteration means the amount of deuterium with respect to hydrogen in molecules. The Big Bang nucleosynthesis theory predicts the initial abundance D/H should be $A_v=1.6 \times 10^{-5}$ at the birth of the Universe. In pre-stellar cores, the dense ($\mathrm{n_H} > 10^5$) and cold ($T<20$K) environments make deuterium fractionation, a process making the increase of the molecular deuteration, occur. Some observations have proved that the ratio of several deuterated molecules, such as $\mathrm{N_2D^+}$, are effective tracers to the star-forming regions. Because the slow reaction rate of deuterium fractionation, it is hard to reach equilibrium state before pre-stellar cores collapse. Measuring the deuterium fraction of species could help us know the ages of prestellar cores. \par

A variety of astrochemical studies have been carried out to model deuterium fractionation. \citet{2004A&A...418.1035W} considered a reduced chemical network, including the spin states of $\mathrm{H_2}$, $\mathrm{H_2^+}$, $\mathrm{H_3^+}$ and $\mathrm{H_2D^+}$. This network assumed heavy elements, like $\mathrm{C}$, $\mathrm{N}$, $\mathrm{O}$, etc, are fully depleted. Extending this work, \citet{Flower2006a}, \citet{2009JChPh.130p4302H}, \citet{Pagani2009} and \citet{2010A&A...509A..98S} included updated reaction rates for spin states and deuterated forms of {\HH} and {$\rm H_3^+$}. \citet{2012A&A...547A..33V} presented networks including molecular species with up to three atoms. \citet{Kong2015} extended these works to include $\mathrm{H_3O^+}$ to acquire more precise results. As a consequence, the abundances of electrons, water, $\mathrm{HCO^+}$, $\mathrm{DCO^+}$, {\NNHp}, {\NNDp} are improved and have a good agreement with the even more extensive network of \citet{Sipila2013}. More recently, \citet{Majumdar2016a}, based on the work of \citet{2015ApJS..217...20W}, presented a complete network including spin state chemistry with 13 elements (\Hy, \He, \C, \N, \Ox, \Si, \So, \Fe, \Na, \Mg, \Cl, \Ph, \F).

The astrochemical models described above have been utilised mostly in single-grid or one-dimensional simulations. However, with this limitation, these simulations could not follow the chemodynamical influence of turbulence, which requires ideally three-dimensional simulations. To investigate such effects, \citet{Goodson2016} developed an approximate model to estimate the levels of [{\NNHp}] and [{\NNDp}], and thus the degree of deuteration. This study used pre-computed tables of chemical evolution of the single-grid models of \citet{Kong2015} to parameterize the growth rate of {\NNHp} and {\NNDp} via the abundances of each of them. With this approximate model, the coupled deuterium fractionation chemistry with 3-D magnetohydrodynamics (MHD) simulations of massive turbulent, magnetized PSCs was studied. However, in addition to its approximate nature, the limitations of this method are the lack of information upon the abundances of other species.

In contrast, the study of \citet{Kortgen2017} followed the dynamical influence on deuterium fractionation by coupling the chemical network of \citet{2004A&A...418.1035W} to PSC hydrodynamic simulations. Recently, the synthetic observations of one selected model from \citet{Kortgen2017} are performed in \citet{2020arXiv201016020Z}. Then, in a follow-up work, the deuterium fractionation of magnetized and turbulent filaments was investigated by \citet{Kortgen2018a}. These studies coupled the deuterium fractionation network to MHD simulations by utilizing the {\KROME} package \citep{Grassi2014a} to solve the chemical reactions in each time step of the simulation. Some limitations of this work include the assumption of fully-depleted heavy elements and the use of threshold-based sink particles to replace high density regions.
% , i.e., as a subgrid model for star formation.
%jct - did they really do this for star formation? Did they have feedback?
%cjhsu - The sink particle is not for star formation. It is a compromise of resolution. Otherwise, they might have some trouble to run the simulation for an enough long time. There is no any feedback. Bovino2019 also has the same problem, but they use a higher threshold n_H=2.7e9 (which was 3e7 in Kortgen2017). 
Furthermore, the increasing size of the chemical reaction network limits the simulation efficiency. In a recent work, \citet{Bovino2019} also applied {\KROME} to investigate deuterium fractionation on multiple scales of clumps and cores. They considered heavy elements, like {\C}, {\N}, and {\Ox} and focused on discussing the influence of time-dependent depletion/desorption reactions to {\CO} depletion structures inside the clumps and cores.

In this paper, we re-examine the work of \citet{Goodson2016} and further discuss the influence of other chemical parameters in the standard, fiducial PSC model of the Turbulent Core Accretion theory of \citet{2003ApJ...585..850M}. We discuss our numerical methods, including choices of initial conditions and chemical modeling, in \S\ref{sec:method}. The results are presented in \S\ref{sec:result}, including the basic results of the fiducial model (\S\ref{sec:fiducial}), the effects of different chemical conditions (\S\ref{sec:chemical}), average core properties and comparison to observed systems (\S\ref{sec:obs}), and core structure, kinematics and dynamics (\S\ref{S:dynamics}). A summary and our conclusions are given in \S\ref{sec:discussion}. 
% We conclude in \S\ref{sec:conclusion}.

\section{Numerical Methods}
\label{sec:method}

We use the adaptive mesh refinement (AMR) code ENZO (v2.5) \citep{Bryan2014} to run our magnetohydrodynamics (MHD) simulations and adopt the MUSCL-Dedner method with Harten-Lax-van Leer (HLL) Riemann solver to solve the ideal MHD equations. The PSC is assumed to be isothermal and collapses under its self-gravity. We set the ratio of specific heats to be $\gamma = C_p/C_{V} = 1.001$ to approximate isothermal conditions, with fiducial temperature of 15~K. With this approximation, the highest density regions reached in the simulations would have a temperature of about 20~K, so the collapse is slightly stabilised by this extra thermal pressure compared to the pure isothermal case. However, since thermal pressure is generally quite small compared to turbulent ram pressure and magnetic pressure, this effect has a very minor influence on the dynamics. We also run the astrochemical evolution at fixed temperature, i.e., 15~K, in the fiducial case. The self-gravity is solved by the fast Fourier transform technique under periodic boundary, which is the same as the boundary condition of the gas.

Higher spatial resolution grids are introduced by a Jeans length refinement criterion, i.e., so that the local Jeans length is resolved by 8 cells. However, note that only up to 5 levels of refinement are introduced, so this Jeans \citep{Truelove1997} criterion is not satisfied at higher densities. The equivalent resolution is $2048^3$ in our settings ($64^3$ in the base level). No sink particles are used in our simulations, so the time step becomes smaller and smaller as time evolves. In practice, this means the simulation can be run for about one free-fall time. We note that fragmentation is suppressed by the presence of magnetic fields in the PSC, together with the assumed initially centrally concentrated density profile.

\subsection{Initial conditions}

We start from the fiducial (S3M2) model described in \citet{Goodson2016}. The model initialises a spherical, turbulent, magnetised PSC according to \citet{2003ApJ...585..850M}. The core has $60\:\Ms$ and its density decreases with radius with a power-law exponent $k_{\rho} = -1.5$, i.e., $\rho \propto r^{-1.5}$. This profile is implemented in an approximate way, described below.
The mass surface density of the surrounding clump environment, which sets the bounding pressure, is set to $\Sigma_{\rm cl} = 0.3\:{\rm g\: cm}^{-2}$, and the relationships described in \citet{2003ApJ...585..850M} give the radius of the PSC to be
\begin{equation}
    R_{c} = 0.057\ \Sigma_{\rm cl}^{-1/2}\ \text{pc} \rightarrow 0.104\ \text{pc}
\end{equation}
and the number density of hydrogen nuclei at the surface to be
\begin{equation}
    n_{\mathrm{H},s} = 1.11 \times 10^6\ \Sigma_{\rm cl}^{3/2} \ \text{cm}^{-3} \rightarrow 1.82 \times 10^5\ \text{cm}^{-3}.
\end{equation}
The average number density in the core is $\bar{n}_{\rm H} = 1.97 \times 10^5 {\rm cm}^{-3}$ and thus the associated free-fall time of the average density is 
\begin{equation}
    \bar{t}_{\text{ff}} = \sqrt{\frac{3\pi}{32G\bar{\rho}}} \rightarrow 76\ \text{kyr}.
\end{equation}

The model assumes the PSC is magnetized by a cylindrically symmetric $B$-field along the z-axis. The field strength is determined by the mass-to-flux ratio normalized with the critical value $M_{\Phi}$ \citep{Mouschovias1976}
%,Krumholz2019}
\begin{equation}
    \mu_{\Phi} = \frac{M}{M_{\Phi}} = \frac{2\pi G^{1/2}M}{\Phi}
\end{equation}
where $\Phi$ is the magnetic flux. The value of $\mu_{\Phi}$ is 2, making the system slightly supercritical. Thus the average $B$-field in the PSC is about 0.44 mG (see below).

\subsubsection{Density profile}
\label{profile}

We use the same density profile described in \citet{Goodson2016} to initialize the PSC:
\begin{equation}
    \rho(r) = \rho_0 + \frac{\rho_c-\rho_0}{1+(r/R_f)^{k_{\rho}}}\left(0.5-0.5\tanh{\left[\frac{r-R_c}{R_s}\right]}\right),
\end{equation}
where $\rho_c$ and $\rho_0$ are the defined central gas density and ambient gas density. The flattening radius, $R_f = 0.15 R_c$, removes the singularity at the origin. The central density is then defined to be $\rho_c = \rho_s [1.0 + (R_c/R_f)^{k_{\rho}}]$, where $\rho_s = n_{{\rm H},s} \mu_{\rm H}$ is the density at the surface and $\mu_{\rm H}$ is the mass per H. 

To balance pressures at the surface of the PSC, we set a density jump $\rho_0 = 0.3 \rho_s$ and give the ambient gas a higher temperature $T_0 = (10/3) T_c \rightarrow 50$~K at the core surface. The $\rm tanh$ function ensures a smooth transition at the surface. In our case, almost all hydrogen exists in molecular form. Therefore, the particle number density at the surface of the core is 
\begin{equation}
    n_{s} \approx 1.82 \times 10^5\ \text{cm}^{-3} \times \frac{\mu_{\Hy}}{\mu} \approx 1.093 \times 10^5\ \text{cm}^{-3},
\end{equation}
where the mean molecular weight $\mu \approx 2.3516\ amu$.
%jct - I think some of this needs to be fixed in terms of n_H and \mu_H. Where does 0.6 come from? We should be assuming 1 He for 10 H.
%cjhsu - Yes, we assume 1 He for 10 H, so the total density is 1.4 * n_H and the mean molecular weight is around 2.3516. Then the number density is around n_H * 1.4 / 2.3516. The latter term (1.4/2.3561) is around 0.6.
Therefore, the values of $\rho_0$, $\rho_s$ and $\rho_c$ are $1.2905 \times 10^{-19}$, $4.30153 \times 10^{-19}$, and $7.82733 \times 10^{-18}\:{\rm g\:cm^{-3}}$.

\subsubsection{Magnetic field profile}

We use the method described in \citet{Goodson2016} to set the magnetic field, which is described by:
\begin{equation}
    B(\xi) = B_0 + \frac{B_c-B_0}{1+(\xi/R_f)^{0.5}}\left(0.5 - 0.5 \tanh\left[\frac{\xi-R_c}{R_s}\right]\right),
\end{equation}
where $\xi = \sqrt{x^2+y^2}$ is the cylindrical radius. The magnetic field is assumed to decrease with cylindrical radius as a power law $B(\xi) \propto (\xi/R_c)^{-0.5}$, but also flatten in the center and taper at the surface, as was implemented for the density profile. The power law keeps the magnetic pressure $(B^2/8\pi) \propto r^{-1}$. Then, $B_0 = B_s$ represents both the ambient magnetic field and the surface magnetic field. The central magnetic field is given by $B_c = B_s[1.0+(R_c/R_f)^{0.5}]$ and the surface field is given by 
\begin{equation}
    B_s = \frac{3G^{1/2}M_c}{2\mu_{\Phi}R_c^2}.
\end{equation}
To obtain $\mu_{\Phi} = 2$ for the PSC, implies that $B_s =0.22$~mG.

% temperature growth(initial/final temperature)
% updated network
% individual time
% advection

\subsubsection{Turbulence}
\label{sec:turbulencediscussion}

We initialize the turbulent velocity field according to a Burgers turbulence power spectrum $E(k) \propto k^{-2}$. 
% The power spectrum is plotted in Fig.~\ref{spectrum}. 
The value of wavenumber $k$ runs through all integers satisfying $\displaystyle 1 < |kL|/(2\pi) < N/2$, where $N$ and $L$ are the number of cells and the length along one edge of the domain. The field is initialized as a solenoidal (divergence-free) field. The code implementation applying \texttt{FFTW} is utilised from the {\GANDALF} code \citep{Hubber2018a}.

The magnitude of the initial turbulent velocity field is scaled to have a one-dimensional velocity dispersion $\sigma = 0.99 \text{km\ s}^{-1}$, which causes the core to be moderately supervirial in the beginning. We calculate the total momentum inside the core and shift the mean velocity to zero to ensure the core stays at the domain center. There is no driven field and the energy of the turbulence dissipates as time evolves.

We note that we use a different power spectrum from \citet{Goodson2016}, which was $E(k) \propto const$, yielding more small scale power. The case of the Burgers turbulent scaling is expected to be more valid in molecular clouds, although there are few constraints on the particular regions that may be massive PSCs. In addition, we note that \citet{Kortgen2017,Kortgen2018a} also used the Burgers power spectrum for their initial conditions. 

\subsection{Chemical model}
\label{chemistry}

The chemical model used in the simulation is updated from \citet{Kong2015}. The new chemical network is composed of the same 132 species and a total of 3466 reactions. For simplicity, we have not considered time-dependent depletion/desorption in our model. This also means we are able to investigate the effect of the level of heavy element depletion in a controlled manner. This approach is justified in the limit that the depletion time ($t_{D}\propto n_{\rm H}^{-1}$) is relatively short compared to the dynamical or free-fall time ($t_{\rm ff}\propto n_{\rm H}^{-1/2}$) of the PSC so that the level of heavy element depletion has approached an approximate equilibrium prior to the collapse phase of the core.

\begin{figure}
    \centering
    \includegraphics[scale=0.45]{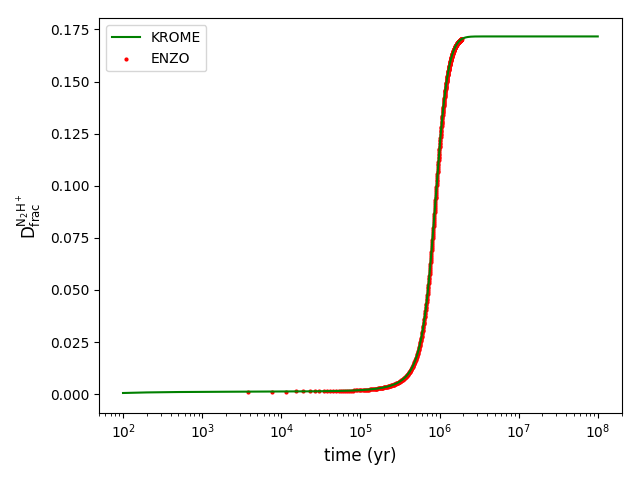}
    \caption{Test case of evolution of $D^{\NNHp}_{\text{frac}}$ with initial {\OPRHH} $= 0.1$. Green line shows the evolution in a single grid run by {\KROME}. Red points shows the result run by  {\ENZO}+{\KROME} in a $4^3$ cube with static uniform gas. The chemical abundance evolution is consistent in the two cases.}
    %does not change by {\ENZO}}
    \label{dfrac}
\end{figure}

The main differences between our network and that of \citet{Kong2015} are minor changes in the dissociative recombination rates of all the forms of $\mathrm{H_3^+}$. These values were interpolated from an external table \citep{Pagani2009} in \citet{Kong2015}. We have updated these values using the analytic expressions of \citet{Majumdar2016a}. After these modifications, there is a 5\% difference in the obtained equilibrium value of \DfracNNH resulting in a standard test case compared to that of \citet{Kong2015}. Figure~\ref{dfrac} demonstrates this result of our updated network. We use the fiducial parameters of \citet{Kong2015} with initial {\OPRHH} $= 0.1$ to do the test. The final equilibrium value of our model is 0.1716, which is close to the value of 0.181 found by \citet{Kong2015}.

We utilize {\KROME} \citep{Grassi2014a} to import the chemistry solver into {\ENZO}. {\KROME} is responsible for calculating the local equilibrium states of species in cells. The advection between cells is handled by {\ENZO}. By default, {\ENZO} calls the chemistry solver at each time step. However, due to the size of the network, we speed up the calculation by only evaluating the chemical evolution at coarser time intervals. Fortunately, the deuterium fractionation reactions proceed much more slowly than the dynamical evolution of the gas. Thus, we are able to run the chemistry solver with a longer time step, i.e., after every 0.01 global initial free-fall times. The chemical model is applied everywhere in the domain, i.e., including regions outside the PSC.

% verify the performance of longer timestep
% In Section ~\ref{profile}, we described there is a jump of density and temperature at the surface of the core. In real world, we expect there is no this kind of jump and the variation of density and chemical abundance would be smooth. To avoid the chemistry reach different equilibrium state inside and outside core, the chemical reactions are forbidden when the temperature is higher than 30 K (since the smooth transition). It means the initial chemical abundances would be the boundary conditions of the core. Also, to avoid the program crashed by some point whose temperature is out of the range of the reactions, we set another temperature threshold 5 K to active the chemistry solver. \par

In our fiducial case, we use the same parameters as in \citet{Kong2015}: gas temperature $\rm T=15K$, cosmic-ray ionization rate $\zeta=2.5 \times 10^{-17}\ \text{s}^{-1}$, heavy element depletion factor \mbox{$f_D = 10$}, ratio to Habing FUV field $G_0=1$, visual extinction $A_V=30$ mag, dust-to-gas mass ratio DGR$=7.09 \times 10^{-3}$, dust particle radius $a_0=1.0 \times 10^{-5}\ \text{cm}$ and dust grain density $\rho_{\text{GRAIN}}=3.0\ \text{g cm}^{-3}$. All elements are assumed to initially be in atomic form, except for ortho/para- {\HH} and {$\rm HD$} (see Table~\ref{tab:initialabund}).

As C and O-bearing molecules have been measured to deplete more efficiently from the gas phase compared to N-bearing ones \citep[e.g.,][]{Caselli2002, Bergin2002}, we use $f_D^{\C, \Ox}$ to represent the depletion factor of {\C} and {\Ox} separately from the other heavy elements, i.e., N, represented by $f_D^N$. We use various values of gas temperature, {\CO} depletion factor, initial {\OPRHH} and cosmic-ray ionization rate in our simulations. Table~\ref{tab:chemmodel} shows the different choices used. 
%The initial abundances of the fiducial model are listed in Table~\ref{tab:initialabund}.  
The abundances of ortho- and para- {\HH} are controlled by {\OPRHH} and the abundances of atomic heavy elements are decided by the depletion factors ($f_D^{\C, \Ox}$ and $f_D^{\N}$). However, we also investigate two models in which the core is chemically aged by running the network at a constant density of $n_{\mathrm{H}}=1.97 \times 10^5\:{\rm cm}^{-3}$ for $3\times 10^5$ and $6\times10^5$~yr. 
%jct - check this
%cjhsu - All heavy elements (C, N, O) are in atomic form in the beginning. I am wondering whether should also describe the OPR at chemical ages = 3e5 / 6e5 yr, which are 1.729e-02 and 2.881e-03.

%jct - I consider this a deficiency in the hydrodynamics, not the chemistry, so it should be reported earlier in the other section.
%A noticeable thing is the gas temperature is fixed in our simulations when we run the chemical routine. This helps us get an ideal isothermal result. 
%By contrast, the isothermal condition in the hydrodynamic part is implemented by setting the specific heat $\gamma = 1.001$. The temperature could increase a little bit up to about $20K$, but the effect should not make huge difference.

\begin{table*}
  \begin{tabular}{lcccccc}
  \hline
  model & {\OPRHH} & T & $\zeta$ & $f_\text{D}^{\N}$ & $f_\text{D}^{\C, \Ox}$ & {\DfracNNH} \\
  \hline
  Fiducial  & 0.1 & 15K & $2.5 \times 10^{-17}\ \text{s}^{-1}$  & 10 & 10 & - \\
  OPR0.01  & 0.01 & 15K & $2.5 \times 10^{-17}\ \text{s}^{-1}$  & 10 & 10 & - \\
  T10 & 0.1 & 10K & $2.5 \times 10^{-17}\ \text{s}^{-1}$  & 10 & 10 & - \\
  CR10x & 0.1 & 15K & $2.5 \times 10^{-16}\ \text{s}^{-1}$  & 10 & 10 & - \\
  fDCO100 & 0.1 & 15K & $2.5 \times 10^{-17}\ \text{s}^{-1}$  & 10 & 100 & - \\
  fDCO1000 & 0.1 & 15K & $2.5 \times 10^{-17}\ \text{s}^{-1}$  & 10 & 1000 & - \\
  Aged3e5 & $1.729 \times 10^{-2}$ & 15K & $2.5 \times 10^{-17}\ \text{s}^{-1}$  & 10 & 10 & $7.58 \times 10^{-3}$ \\
  Aged6e5 & $2.881 \times 10^{-3}$ & 15K & $2.5 \times 10^{-17}\ \text{s}^{-1}$  & 10 & 10 & $5.38 \times 10^{-2}$ \\
%   OPR3 & 3 & 15K & $2.5 \times 10^{-17}\ \text{s}^{-1}$  & 10 & 10 & - \\
%   OPR0.001 & 0.001 & 15K & $2.5 \times 10^{-17}\ \text{s}^{-1}$  & 10 & 10 & - \\
  OPR0.01\_fDCO1000 & 0.01 & 15K & $2.5 \times 10^{-17}\ \text{s}^{-1}$  & 10 & 1000 & - \\
  OPR0.001\_fDCO1000 & 0.001 & 15K & $2.5 \times 10^{-17}\ \text{s}^{-1}$  & 10 & 1000 & - \\
  % new models to be added
  fDN100\_fDCO1000 & 0.1 & 15K & $2.5 \times 10^{-17}\ \text{s}^{-1}$  & 100 & 1000 & - \\
  fDN1000\_fDCO1000 & 0.1 & 15K & $2.5 \times 10^{-17}\ \text{s}^{-1}$  & 1000 & 1000 & - \\
  CR4x\_fDCO1000 & 0.1 & 15K & $1.0 \times 10^{-16}\ \text{s}^{-1}$  & 10 & 1000 & - \\
  CR4x\_fDN100\_fDCO1000 & 0.1 & 15K & $1.0 \times 10^{-16}\ \text{s}^{-1}$  & 100 & 1000 & - \\
  \hline
  \end{tabular}
  \caption{List of PSC simulations and their parameters. We first try modifying one parameter at a time compared to the fiducial model to understand its influence, and then test the combination of different parameters. The parameters that are varied are initial {\OPRHH}, temperature ($T$), cosmic-ray ionization rate ($\zeta$), depletion factor of {\C} and {\Ox} ($f_D^{\C, \Ox}$), and depletion factor of N ($f_D^N$). The "Aged" models use the same initial parameters as the fiducial model, but the chemistry has been evolved first at constant density for $3 \times 10^5$ and $6 \times 10^5$ years before the core starts to collapse. We note the values of {\OPRHH} and {\DfracNNH} at the beginning of hydrodynamic simulations in the table.}
  \label{tab:chemmodel}
\end{table*}

\begin{table}
    \centering
    \begin{tabular}{lc}
    \hline
        Species & Abundance ($n_{\rm species}/n_{\rm H}$) \\
    \hline
        p-{\HH} & 5/11 = 0.455 \\
        o-{\HH} & 5/110 = 0.0455 \\
        {\rm HD} & $1.5 \times 10^{-5}$ \\ 
        \He & $1.0 \times 10^{-1}$ \\ 
        \N  & $2.1 \times 10^{-6}$ \\ 
        \Ox  & $1.8 \times 10^{-5}$ \\ 
        \C  & $7.3 \times 10^{-6}$ \\ 
        GRAIN0  & $1.32 \times 10^{-12}$ \\ 
    \hline
    \end{tabular}
    \caption{The initial abundances of the fiducial model. The depletion factor $= 10$ reduces the abundances of {\C, \Ox} and {\N} by a factor of 10 from the default Solar metallicity ISM values.}
    \label{tab:initialabund}
\end{table}

\section{Results}
\label{sec:result}

We evolve each model listed in Table~\ref{tab:chemmodel} for 80\% of the initial mean free-fall time of 76~kyr, i.e., for 61~kyr. 
%jct - check above; but also I think it should be 0.8 t_ff; If we have results to 1 t_ff, then we should show them, I think, e.g., in Fig. 2.
At the last time step, the number density $n_{\rm H}$ reaches $>10^{11}\: {\rm cm}^{-3}$. Simulations are not run for longer times, given that we do not include any sub-grid model for the formation of a protostar and any feedback that it would generate. Since all simulations start from identical physical initial conditions (except for the 10~K case), they should have the same density, temperature, velocity distribution, etc, at the end. However, the differences in investigated model parameters cause them to have different chemical compositions. First we focus on comparing the chemical compositions and related derived properties, especially {\DfracNNH}. Then we examine kinematic and dynamic properties of the core, especially as traced by {\NNDp}.

\subsection{Fiducial model}\label{sec:fiducial}

\begin{figure*}
    % \centering
    \includegraphics[scale=0.85]{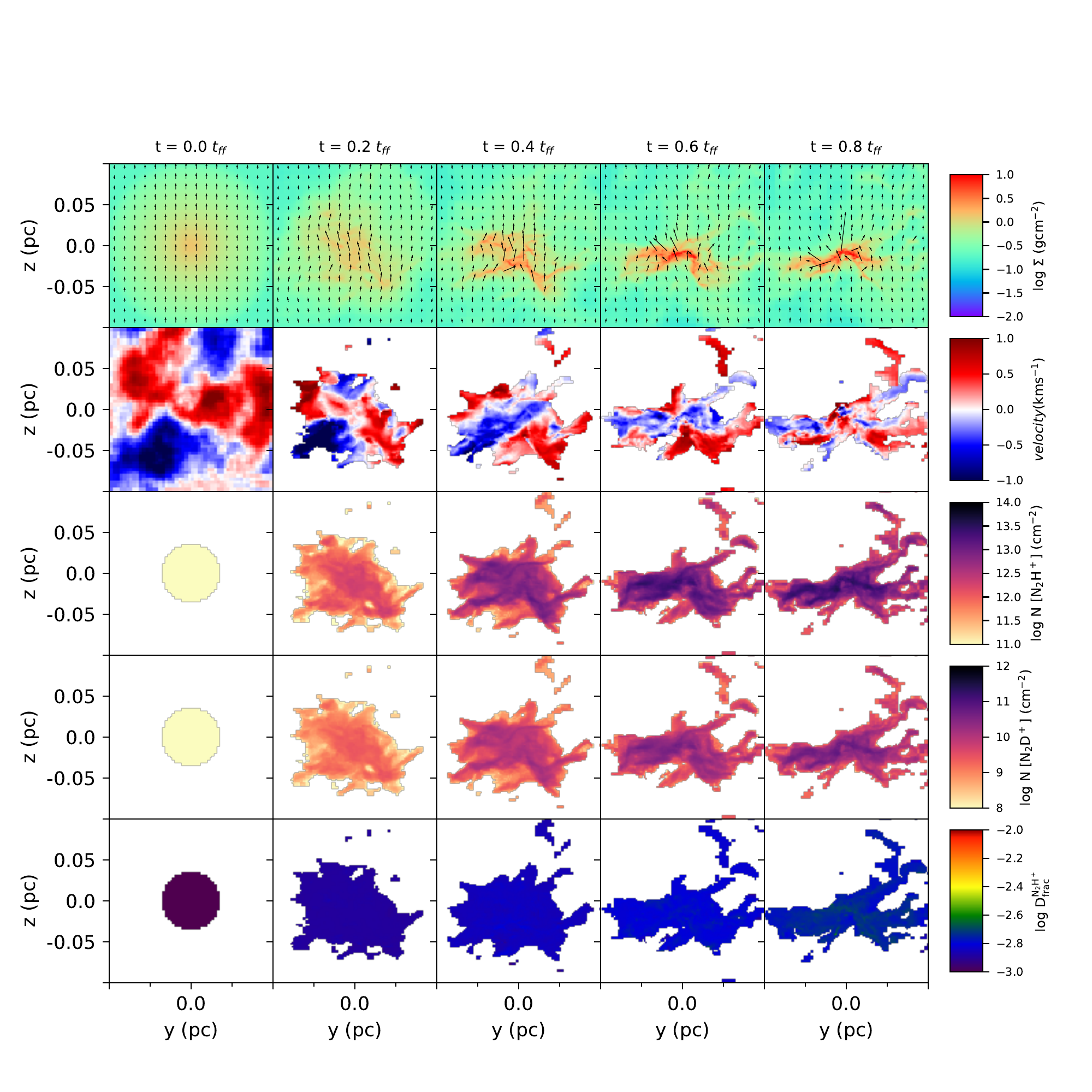}
    \caption{Time evolution of the fiducial model. The columns from left to right show the snapshots at times = 0, 0.2, 0.4, 0.6 and 0.8 of the initial core mean free-fall time of 76 kyr. Each row shows properties projected along the $x$-axis. From top to bottom these are: mass surface density; line-of-sight velocity weighted by {\NNDp} (except at $t=0$, when it is weighted by $\Sigma$) (see text); column density of {\NNHp}; column density of {\NNDp}; and the deuterium fraction, {\DfracNNH}. The lines in the first row indicates the average (mass-weighted) projected magnetic field orientations.}
\label{fiducial_x}
\end{figure*}

\begin{figure*}
    % \centering
    \includegraphics[scale=0.85]{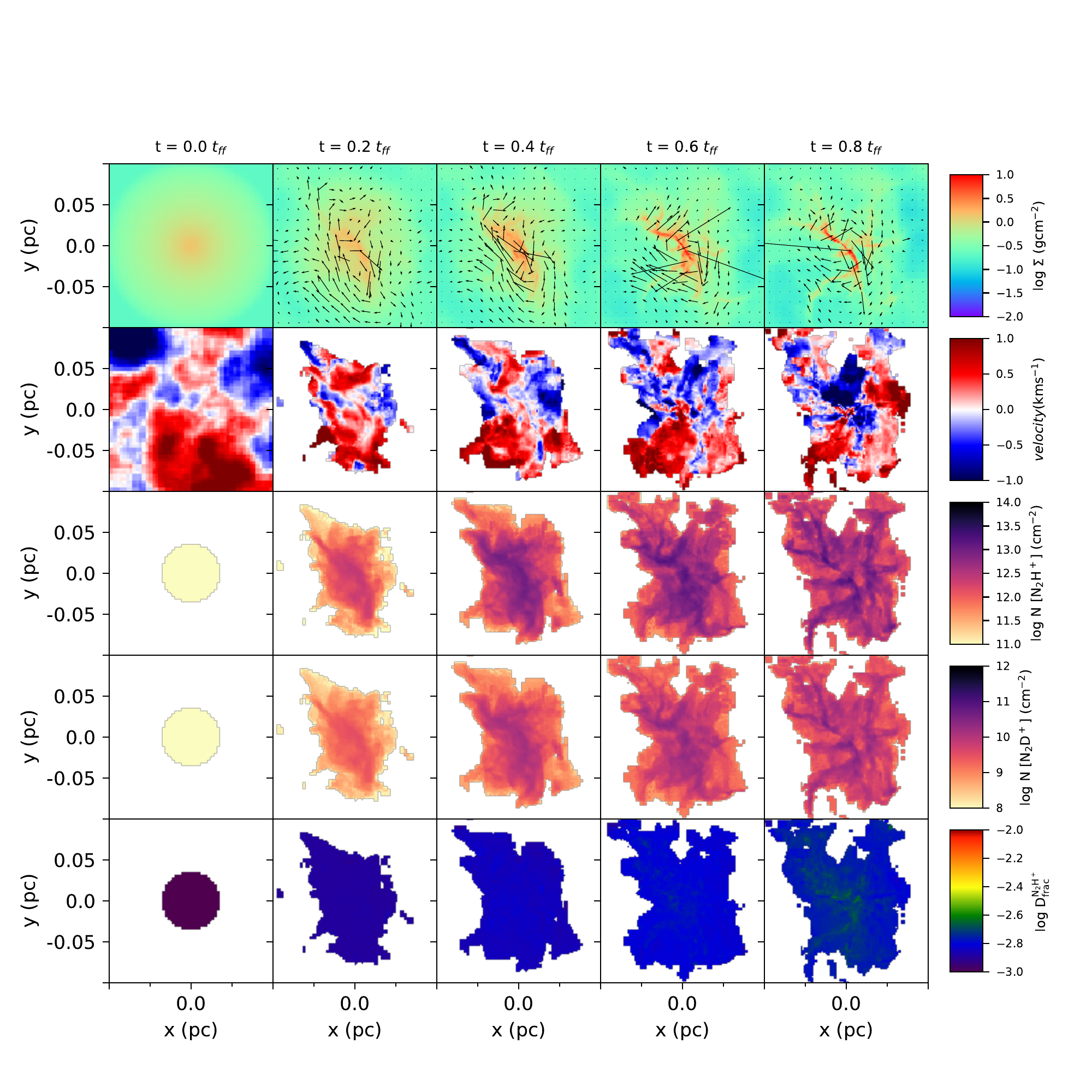}
    \caption{The same as Figure~\ref{fiducial_x}, but now with the panels showing quantities projected along the $z$-axis.}
\label{fiducial_z}
\end{figure*}

Figures~\ref{fiducial_x} and \ref{fiducial_z} show the results of the fiducial simulation, with views of the core projected along the $x$-axis and $z$-axis, respectively. In these figures, the columns from left to right show outputs at 0, 0.2, 0.4, 0.6 and 0.8 $t_{\rm ff}$. The rows from top to bottom are the mass surface density, radial velocity weighted by {\NNDp} abundance, column density of {\NNHp}, column density of {\NNDp} and deuterium fraction of {\NNHp}. To compare with observational data, we follow the criteria in \citet{Goodson2016}, setting a minimum density threshold $n_{\rm H} = 8 \times 10^5\: {\rm cm}^{-3}$, corresponding to roughly 10\% of the critical density of the (3-2) transitions \citep{2013A&A...555A..41M}, when we calculate the column density of {\NNHp} and {\NNDp}. The effect of whether or not to impose this threshold is examined later.

We see from Figures~\ref{fiducial_x} and \ref{fiducial_z} that the core collapses mostly along the large scale $B$-field direction, i.e., the $z$-axis, to form a flattened and eventually filamentary structure. When viewed along the $x$-axis (Figure~\ref{fiducial_x}), the maximum mass surface density grows from around 0.18 to 4.1 $\rm g\ cm^{-2}$ and the magnetic field remains roughly perpendicular to the filament. The average line-of-sight velocities show large-scale structures, dominated at first by the large scale modes of the turbulent initial conditions. Note that at $t = 0$ in Figs.~\ref{fiducial_x} and \ref{fiducial_z}, we have weighted the velocities simply by density because no {\NNDp} has yet formed, while at later times they are weighted by {\NNDp} abundance. 
%Note that this velocity structure is quite different from the simulation presented by \citet{Goodson2016}. 
%jct not sure we need this comment here:
%We discuss the turbulence further in Section~\ref{sec:turbulencediscussion}. 
At the end of the simulation, the line-of-sight velocity field shows a high dispersion near the core center, with values ranging from $-2.0$ to $+3.6$ km/s. The overall velocity dispersion is discussed in more detail in \S\ref{S:dynamics} in relation to the dynamical state of the core.
%jct - these numbers seem to be outside the range of the scalebar... is the scalebar really log v?
%cjhsu - The range is from Goodson et al. 2016. I use the same colormap to easier compare the results.
%jct - we should discuss the velocity dispersion of the core at some point, like we did in Goodson's paper... and the virial parameter

The column densities of {\NNHp} and {\NNDp} grow as time evolves.
%and {\DfracNNH} grow as the time evolves. 
The peak value of $N_{\rm N_2H^+}$ increases to more than $10^{14}\ {\rm cm}^{-2}$, while that of {\NNDp} reaches about $10^{11}\ {\rm cm}^{-2}$. However, note that these values apply to very localised regions within the core, with average values being much lower.
%jct - I don't know what the next statement means
%cjhsu - I try to explain the high column density only appears in a tiny area (could be 0.001 pc * 0.001 pc), which is much smaller than the beam size of the observational data that we are using to do comparison in the end. 
%However, these values are from one grid. They may not be able to represent the real column density of the core. 
%We will discuss the column density of the core later. 
The value of {\DfracNNH} also increases with time,
%with {\NNHp} and {\NNDp} but the rate is relatively slow. In the end, the 
reaching a peak value of about $2 \times 10^{-3}$.

In Figure~\ref{fiducial_z}, the core is viewed projected along the $z$-axis, which is the orientation direction of the initial large-scale, cylindrically symmetry magnetic field. Comparing with Figure~\ref{fiducial_x}, the spatial structure of the core is now somewhat more dispersed, although the mass surface density reaches similar peak values. The magnetic field orientations that arise in this plane are more random, since they lack the component of the original large-scale field and arise solely from the motions due to initial turbulence and gravitational collapse that drag the $B$-field with them. The line-of-sight velocities also reach similar local dispersions around the core center as were seen in the $x$-axis projection. The column densities of {\NNHp} and {\NNDp} and their ratio, {\DfracNNH}, also reach similar values as seen in the previous view. We conclude that the large-scale magnetic field does not significantly change the patterns of most of these results. In particular, the projections along the x-axis are sufficient for us to analyze the astrochemical results.

\begin{figure*}
    % \centering
    \includegraphics[width=\linewidth]{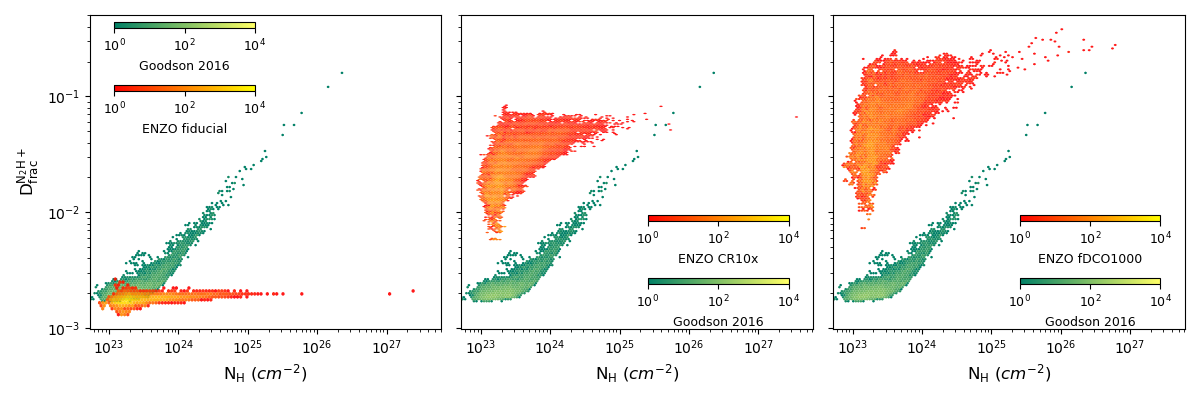}
    \caption{Scatter plot of total H column density versus {\DfracNNH}. In each panel, the green-yellow colour-map shows the fiducial model of \citet{Goodson2016} at the end of the evolution, i.e., at 0.8~$t_{\rm ff}$. From left to right, we plot three of our models in red-yellow colour-map: fiducial, CR10x and fDCO1000 (see text and Table~\ref{tab:chemmodel}).}
\label{coldens-dfrac}
\end{figure*}

In our fiducial model, we see that the growth of {\DfracNNH} is quite slow so that after 80\% of an initial free-fall time the peak values are only $\sim0.002$, which is smaller than seen in many observed PSCs (see \S\ref{S:introduction}). In contrast, \citet{Goodson2016} reported a peak value of {\DfracNNH} about 50 times higher than ours at the equivalent time. In Figure~\ref{coldens-dfrac}, we plot the deuterium fraction versus column density of all the cells in our simulations and compare to the results of \citet{Goodson2016}. The left panel shows the result of the fiducial model, which indicates that the differences arise mostly from the high column density regions. Our derived values of {\DfracNNH} reach a near constant maximum level and do not rise with increasing column density, which is in contrast to the results of \citet{Goodson2016}. We discuss the reasons for this different behaviour in 
%jctfinal - this reference does not appear properly - commenting out
%\S\ref{overestimateissue}
Appendix A, where we explain it as being due to deficiencies in the approximate method of \citet{Goodson2016}.

In summary, our fiducial physical and chemical model cannot achieve the high values of {\DfracNNH} $\gtrsim0.1$ that are observed in many PSCs. One possible explanation is that the physical model is wrong: i.e., the collapse is occurring too quickly and should rather take place over several or many free-fall times \citep[e.g.,][]{Kong2015,Kong2016}. Another possible explanation is that the chemical model is incorrect and that different chemical conditions can allow higher values of {\DfracNNH} to be achieved quickly within about one free-fall time. This second possibility is investigated next in \S\ref{sec:chemical}.

% \begin{figure}
%     % \centering
%     \includegraphics[scale=0.5]{fig/scatter2d_nh_x_Dfrac_t8.png}
%     \caption{The scatter plot of column density versus deuterium fraction}
% \label{coldens-dfrac}
% \end{figure}

\subsection{Effect of chemical conditions on abundances of {\NNHp} and {\NNDp}}\label{sec:chemical}

High values of {\DfracNNH}, up to $\sim0.5$, have been observed in several star-forming regions that are thought to harbour massive PSCs \citep[e.g.,][]{Miettinen2012,Kong2016}. 
It is clear that our fiducial model cannot reach such high values in one free-fall time. In the following subsections, we discuss some possible ways to reach higher values of {\DfracNNH} in our simulated PSC. In particular, we investigate the effect of several model parameters, as summarised in Table~\ref{tab:chemmodel}. 
%jct - commenting out - I think Goodson did investigate some variation of parameters.
% cjhsu - He did investigation for initial OPR and chemical ages, but he did not discuss the influence of CR or depletion
%These parameters were also explored in the single grid models of \citet{Kong2015}, but were not considered in the study of \citet{Goodson2016}. 
%Here we review their influences in three-dimensional dynamical environments. 
Our overall goal is to see under what conditions very high values of {\DfracNNH}, similar to those of some observed PSCs, can be achieved in the context of the presented physical model. If such high values of {\DfracNNH} are not achievable, then this would strengthen the case that a different physical model is needed, e.g., one involving less rapid collapse of the PSC via increased magnetic field support, as has been discussed by, e.g., \citet{Kong2016}.

\begin{figure*}
    % \centering
    \includegraphics[scale=0.85]{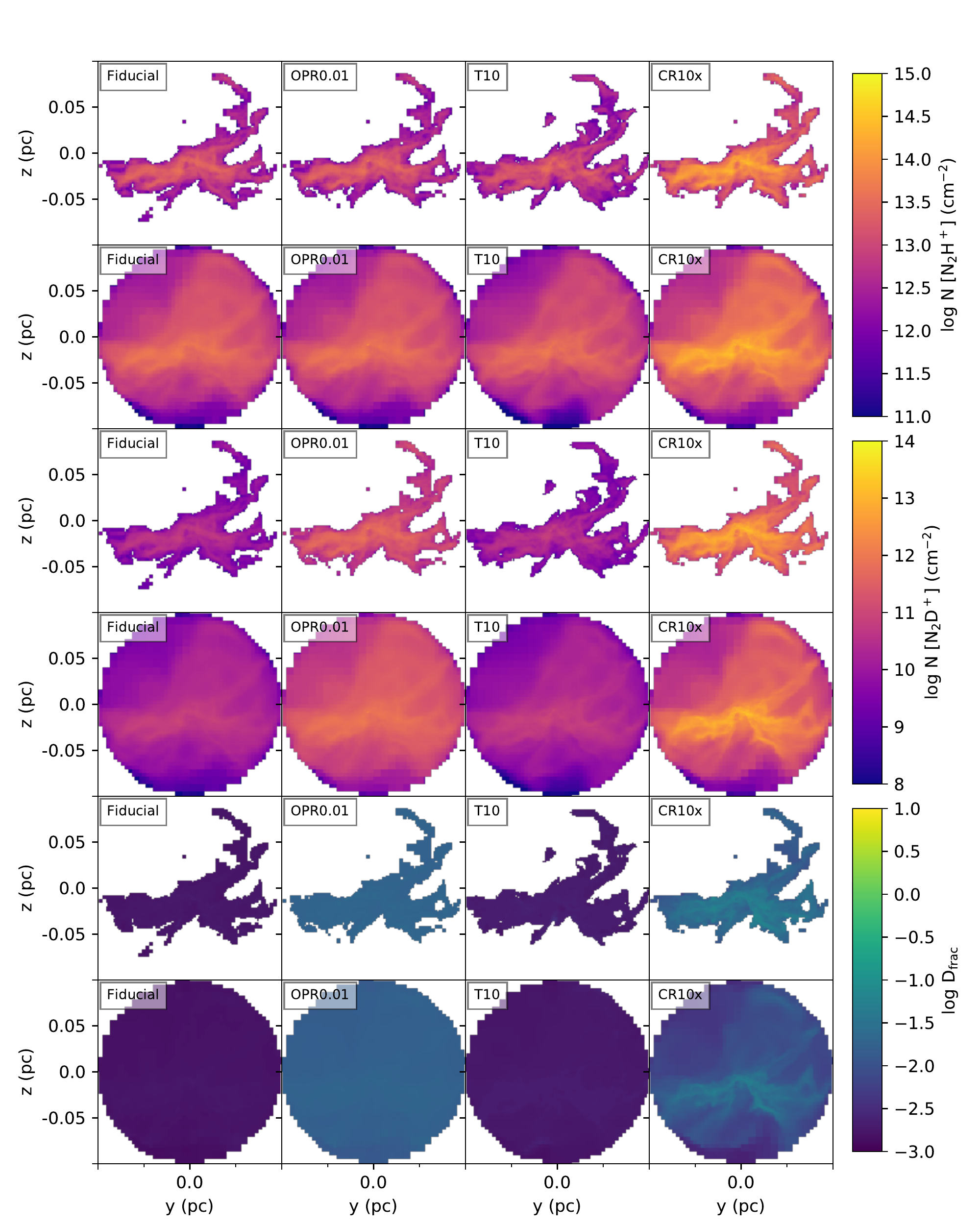}
    \caption{The projections of {\NNHp}, {\NNDp}, and {\DfracNNH} along the $x$-axis at 0.8 $t_{\rm ff}$. These quantities are plotted in two cases (with density threshold and without density threshold) to display the general concentration properties (see text). The ambient gas is subtracted by limiting the volume inside 0.1~pc sphere. Each column represents a different model (left to right): fiducial; OPR0.01; T10; CR10x, with parameters of these models listed in Table~\ref{tab:chemmodel}. The remaining models are plotted in Figure~\ref{fig:dfracmap2-x} and~\ref{fig:dfracmap3_x}}
\label{fig:dfracmap1-x}
\end{figure*}

\begin{figure*}
    % \centering
    \includegraphics[scale=0.85]{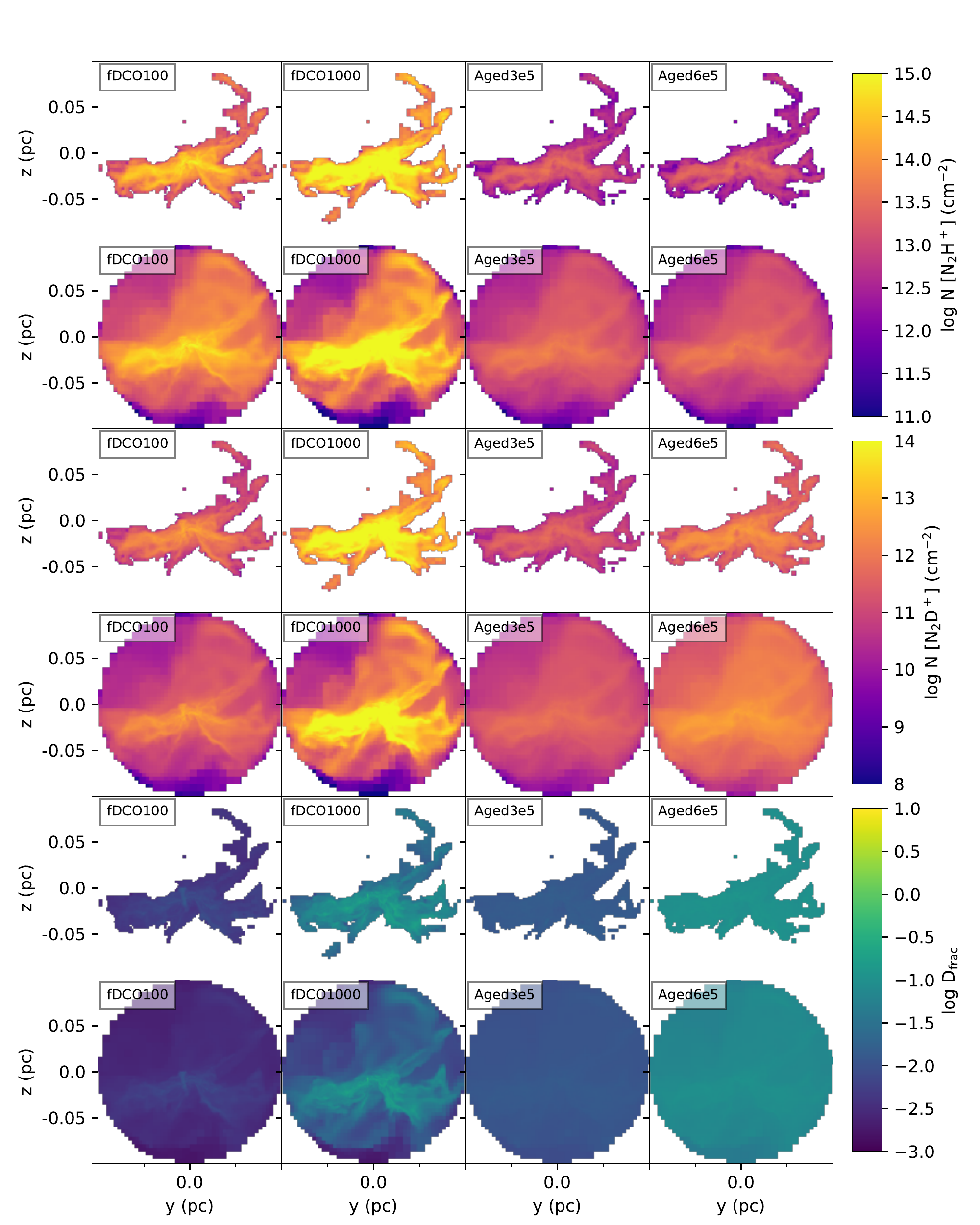}
    \caption{Same as Figure~\ref{fig:dfracmap1-x}, but now for models (left to right): fDCO100; fDCO1000; Aged3e5; Aged6e5, with their parameters listed in Table~\ref{tab:chemmodel}. }
\label{fig:dfracmap2-x}
\end{figure*}

Maps of {\NNHp} and {\NNDp} column densities and their ratio, {\DfracNNH}, projected along the $x$-axis, of the various investigated models at 0.8 $t_{\rm ff}$ are plotted in Figures~\ref{fig:dfracmap1-x} and \ref{fig:dfracmap2-x}. We show versions of the maps (first, third and fifth rows) in which only contributions from cells with number densities $> 8 \times 10^5\ {\rm cm}^{-3}$, corresponding to about 10\% of the critical density of the (3-2) rotational transitions, are counted. However, to investigate the effect of this choice, we also show versions of the maps (second, fourth and sixth rows) without use of this threshold.

% \begin{figure*}
%     % \centering
%     \includegraphics[scale=0.85]{fig/sixPanelz.pdf}
%     \caption{The projection of {\DfracNNH} along z-axis at 0.8 $t_{ff}$}
% \label{dfracComp-z}
% \end{figure*}

% \begin{figure*}
%     % \centering
%     \includegraphics[scale=0.85]{fig/sixPanelx_noThres.pdf}
%     \caption{The projection of {\DfracNNH} along x-axis at 0.8 $t_{ff}$. No density threshold was used when plotting the figure.}
% \label{dfracComp-x-nothres}
% \end{figure*}

%jct - I think we need to see the maps of [N2H+] and [N2D+] column densities in Fig. 5, so that would be 4 more rows for each case. So 6 rows in total for each case. Thus we can split the Fig. into two. (and same for Fig. 8). This is important to see the general concentration of of these species, rather than just their ratios.

\subsubsection{Initial {\OPRHH}}
\label{Ioprh2}

The {\OPRHH} is expected to decline with time in molecular clouds \citep{Flower2006a}. The higher energy of ortho-{\HH} suppresses formation of $\rm H_2D^+$ and thus {\NNDp} and so limits the growth of {\DfracNNH} \citep[e.g.,][]{Pagani1992, Gerlich2002, Walmsley2004, Sipila2010}. %\citep[e.g.,][]{2014prpl.conf..859C}. 
To investigate the effect of initial {\OPRHH} compared to the fiducial model, we use a value that is ten times smaller, i.e., {\OPRHH} $= 0.01$, in the OPR0.01 model. In Figure~\ref{fig:dfracmap1-x}, we see this change causes the deuterium fraction to increase by about one order of magnitude at all locations, with little dependence on column density. Thus reducing the initial {\OPRHH} level is a way to boost {\DfracNNH} in the core. However, we note that in our models this does not lead to a strong spatial concentration of {\DfracNNH} associated with high column density regions. Such a concentration would likely require a similar spatial concentration of {\OPRHH} in the core region. We also note that achieving a low value of {\OPRHH} would require "chemical aging" of the core over a timescale that is expected to be longer than the initial free-fall time of our model \citep[e.g.,][]{Kong2015}, and that this should be considered together with the evolution of other chemical species. Such models are presented below in \S\ref{S:aged}.

\subsubsection{Temperature}

\citet{Kong2015} showed that for temperatures below $\sim20$~K, there is little variation of the growth timescales and equilibrium levels of {\DfracNNH}. However, to check this and to also examine any effects on the dynamical evolution, e.g., affecting fragmentation, we run another isothermal PSC at 10~K in the T10 model. In Figure~\ref{fig:dfracmap1-x}, we see that the {\DfracNNH} map of the T10 model shows a slightly wider distribution than the fiducial model. However, as expected, {\DfracNNH} is at a similar level as the fiducial model. Furthermore, the physical structures that form during the collapse are broadly similar, indicating that thermal pressure support plays a relatively minor role compared to magnetic fields and turbulence.

\subsubsection{Cosmic-ray ionization rate}
\label{sssec:crir}

The cosmic-ray ionization rate, $\zeta$, influences {\DfracNNH} significantly in two aspects: the growth timescale and the final equilibrium value. The timescale to reach the equilibrium state becomes $\sim$100 times shorter and the equilibrium value of {\DfracNNH} $\sim$6 times smaller as $\zeta$ increases from $10^{-18}$ to $10^{-15}\:{\rm s}^{-1}$ \citep{Kong2015}. In the CR10x model, we try a cosmic-ray ionization rate of $2.5 \times 10^{-16}\:{\rm s}^{-1}$, i.e., $10\times$ higher than that of the fiducial model and in line with the value measured in diffuse clouds \citep[e.g.,][]{Indriolo2012}. The result shows that {\DfracNNH} not only increases globally (since we are in the limit where the equilibrium value has not been reached), but has higher values along the dense filaments. In other words, it shows a more concentrated spatial structure. We can see the same phenomenon by comparing with the map plotted without the density threshold. To verify this, we plot {\DfracNNH} versus column density in the middle panel of Figure~\ref{coldens-dfrac}. We see that {\DfracNNH} is increasing rapidly until $N_{\rm H} \sim 3\times 10^{23} {\rm cm}^{-2}$, after which it stays quite constant. Thus we see that higher cosmic-ray ionization rates enhance the influence of density and cause {\DfracNNH} to be a better tracer of high-density regions.

As a caveat, we note that in actual molecular clouds, where cosmic rays are mostly propagating in from the surroundings, we expect higher density regions to have lower cosmic-ray ionization rates, due to shielding by surrounding material and magnetic fields. Still, \citet[][see their figure F.1]{Padovani2018}, predict values close to 10$^{-16}\:{\rm s}^{-1}$ for H$_2$ column densities of about 10$^{23}\:{\rm cm}^{-2}$, including the effects of attenuation. Further independent constraints on the levels of the cosmic-ray ionization rate are needed when making detailed comparisons of our models with observed sources.

\subsubsection{$\mathrm{CO}$ depletion}
\label{codepletion}

As {\CO} destroys $\mathrm{H}^+_3$ and its deuterated forms, a higher level of {\CO} depletion is beneficial to deuterium fractionation \citep{1987IAUS..120..109D, Crapsi2005}. %\citep[e.g.,][]{2014prpl.conf..859C}. 
The level of {\CO} depletion itself also depends on density, temperature, cosmic-ray ionization rate, etc., and will evolve in time towards an approximate equilibrium level. However, for simplicity, we do not include freeze-out reactions and ice phase species in our network. Instead, we model {\CO} depletion as a fixed input parameter, following the treatment of \citet{Kong2015}. The level of depletion is parameterized by the {\CO} depletion factor, $f_D^{\C,\Ox}$, which is the ratio of expected CO abundance if all C is in the form of gas-phase CO and the actual gas phase CO abundance. This method is expected to be approximately valid in cold ($T\lesssim20\:$K), dense ($n_{\rm H}\gtrsim 10^4\:{\rm cm}^{-3}$) protocluster clump environments in which the freeze-out time is short compared to the free-fall time \citep{Walmsley1991}. Note that in our physical model the surrounding clump gas has a density of $n_{\rm H,cl}=6\times10^4\:{\rm cm}^{-3}$.

With reference to the fiducial model, which has $f_D^{\C,\Ox}$ =10, we explore the influence of two higher values of $f_D^{\C,\Ox} =$100 and 1000 in the fDCO100 and fDCO1000 models, respectively. In Figure~\ref{fig:dfracmap2-x}, both models show more concentrated spatial structures. In the densest regions, the 10 times higher depletion factor in the fDCO100 model causes {\DfracNNH} to become about 10 times higher than in the fiducial case. Similarly, the 100 times higher depletion factor in the fDCO1000 model causes {\DfracNNH} to increase by about two orders of magnitude, i.e., reaching values $>$0.1 in less than one free-fall time. 

The relationship between {\DfracNNH} and column density of the fDCO1000 model is also plotted in Figure~\ref{coldens-dfrac} (right panel). It shows the fDCO1000 model has an overall higher level of {\DfracNNH} than the other models so far considered.

%jct - caselli's comment: (some kind of summary of observed values of CO depletion factor).
%maybe refer to observational studies of CO depletion in high-mass star forming regions (Hernandez et al. 2011, Fontani et al. 2012, Giannetti et al. 2014) to suggest which value of CO depletion sounds more appropriate. You can also mention that in local pre-stellar cores, the level of CO depletion in the central regions is predicted to be very large (between 100 and 1000). 

\subsubsection{Initial chemical age}\label{S:aged}

We also investigate starting the simulation from an older chemical age. To obtain chemical abundances at a specific chemical age, we first calculate the mean density of the PSC to be $\bar{n}_{\rm H}=1.97 \times 10^5 \rm cm^{-3}$. We use this mean density and other fiducial model parameters of $T$ and $\zeta$ to evolve a single grid chemical model to obtain abundances of species at a given chemical age. The effect is similar to using a smaller initial {\OPRHH}, but now the chemical composition is no longer starting from pure elemental abundances (of the heavy elements).

\begin{figure}
  \centering
  \includegraphics[scale=0.5]{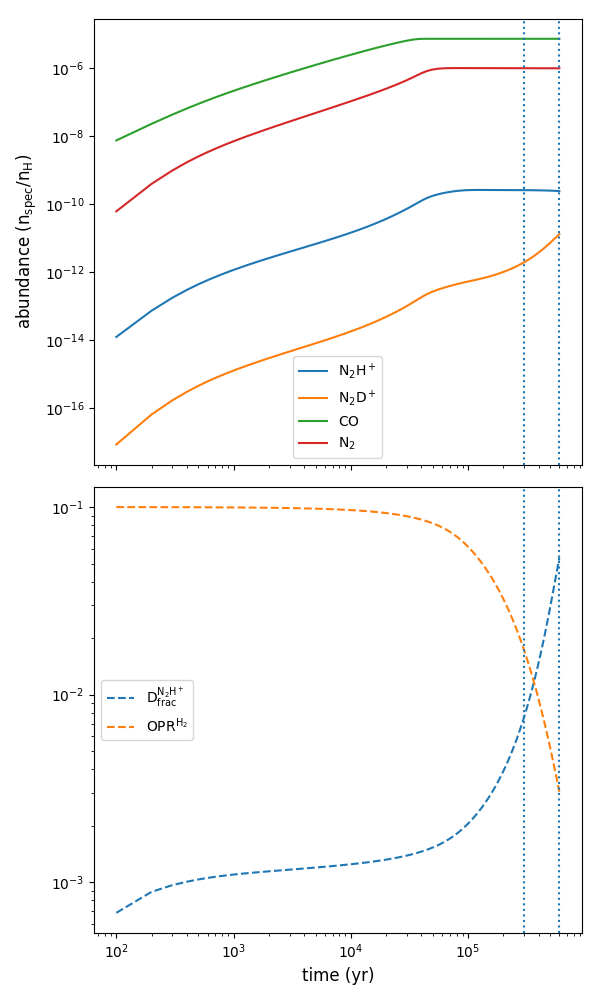}
  \caption{The time evolution of key species ({\CO}, {$\rm N_2$}, {\NNHp}, {\NNDp}) (top panel) and {\OPRHH} and {\DfracNNH} (bottom panel) in the prior evolutionary phase of the PSC. Vertical lines indicate times at $3 \times 10^5$ years and $6 \times 10^5$ years.}
\label{fig:clumpchem}
\end{figure}

We investigate two different chemical ages: $3 \times 10^5$ years and $6 \times 10^5$ years. Figure~\ref{fig:clumpchem} shows the time evolution of key species in this PSC prior evolutionary phase. We see that by $3 \times 10^5$ years, equilibrium levels of CO, {$\rm N_2$} and {\NNHp} have been achieved, while {\NNDp} is still growing. This is because, at this time, {\OPRHH} is undergoing a gradual decline. The values of {\OPRHH} at (3 and 6)~$\times10^5\:$yr are (1.73 and 0.29)~$\times10^{-2}$, while the values of {\DfracNNH} are (0.76 and 5.4)~$\times10^{-2}$, respectively (see Table~\ref{tab:chemmodel}).

In Figure~\ref{fig:dfracmap2-x}, we see that the models Aged3e5 and Aged6e5 that have this described prior chemical evolution are able to reach higher values of {\DfracNNH}. Especially model Aged6e5 can reach {\DfracNNH} $\sim 0.1$, similar to the levels achieved in model fDCO1000. However, now the effect is more global, i.e., there is not as strong a spatial concentration of {\DfracNNH} following the column density structure. We also see that the similar levels of {\DfracNNH} that arise in these models are achieved with quite different absolute abundances of {\NNHp} and {\NNDp}. We will return to these metrics as a way to distinguish between models in \S\ref{sec:obs}.

\subsubsection{Combination of initial {\OPRHH} and {\CO} Depletion}

In \S\ref{Ioprh2}, we saw that lower initial {\OPRHH} helps the growth of {\DfracNNH}, while its spatial structure is not enhanced with density. In contrast, the models of \S\ref{codepletion} with a higher {\CO} depletion factor have a clearer structure of \DfracNNH following column density. Therefore, we try a series of combinations of initial {\OPRHH} and {\CO} depletion to survey whether {\DfracNNH} can have a stronger correlation with column density. We consider two possible {\CO} depletion factors: 10 and 1000, and four possible values of initial {\OPRHH}: 3, 0.1, 0.01 and 0.001. 
% The results are shown in Fig.~\ref{fig:dfracmap_oprco1_x} and Fig.~\ref{fig:dfracmap_oprco2_x}. 
From the results of these eight models, initial {\OPRHH} does not appear to enhance the spatial structure of \DfracNNH. 

Although the initial {\OPRHH} does not further enhance the spatial structure, the models with high {\CO} depletion factor and low initial {\OPRHH} still show the possibility to obtain high values of {\DfracNNH}. We thus focus on two models, OPR0.01\_fDCO1000 and OPR0.001\_fDCO1000, which are listed in Table~\ref{tab:chemmodel} and which will be considered further in the comparison to observed systems (see below).

% cjhsu - I am wondering whether these figures should be added. They may provide little information.
% \begin{figure*}
%     % \centering
%     \includegraphics[scale=0.8]{fig/deualtDecomap1_x.pdf}
%     \caption{Same as Fig.~\ref{fig:dfracmap1-x}, but now for models with initial {\OPRHH} = 3, 0.1, and 0.001 (columns from left to right, respectively), all with {\CO} depletion factor = 10. Note, we use a different colormap scale in the case initial {\OPRHH} = 0.001, given its high level of {\DfracNNH}.}
% \label{fig:dfracmap_oprco1_x}
% \end{figure*}

% \begin{figure*}
%     % \centering
%     \includegraphics[scale=0.8]{fig/deualtDecomap2_x.pdf}
%     \caption{Same as Fig.~\ref{fig:dfracmap1-x}, but now for models with initial {\OPRHH} = 3, 0.1, and 0.001 (columns from left to right, respectively), all with {\CO} depletion factor = 1000. Note, we use a different colormap scale in the case initial {\OPRHH} = 0.001, given its high level of {\DfracNNH}.}
% \label{fig:dfracmap_oprco2_x}
% \end{figure*}

\subsubsection{Combination of {\N} and {\CO} depletion factors}

In \S\ref{codepletion}, we found that increased {\CO} depletion factor enhances the spatial concentration of {\DfracNNH}. In reality, {\N} depletion should also happen where {\CO} depletion occurs, although perhaps at a more modest level \citep{Flower2006b}. The level of N depletion will naturally affect the absolute abundances of {\NNHp} and {\NNDp}.

\begin{figure*}
    % \centering
    \includegraphics[scale=0.85]{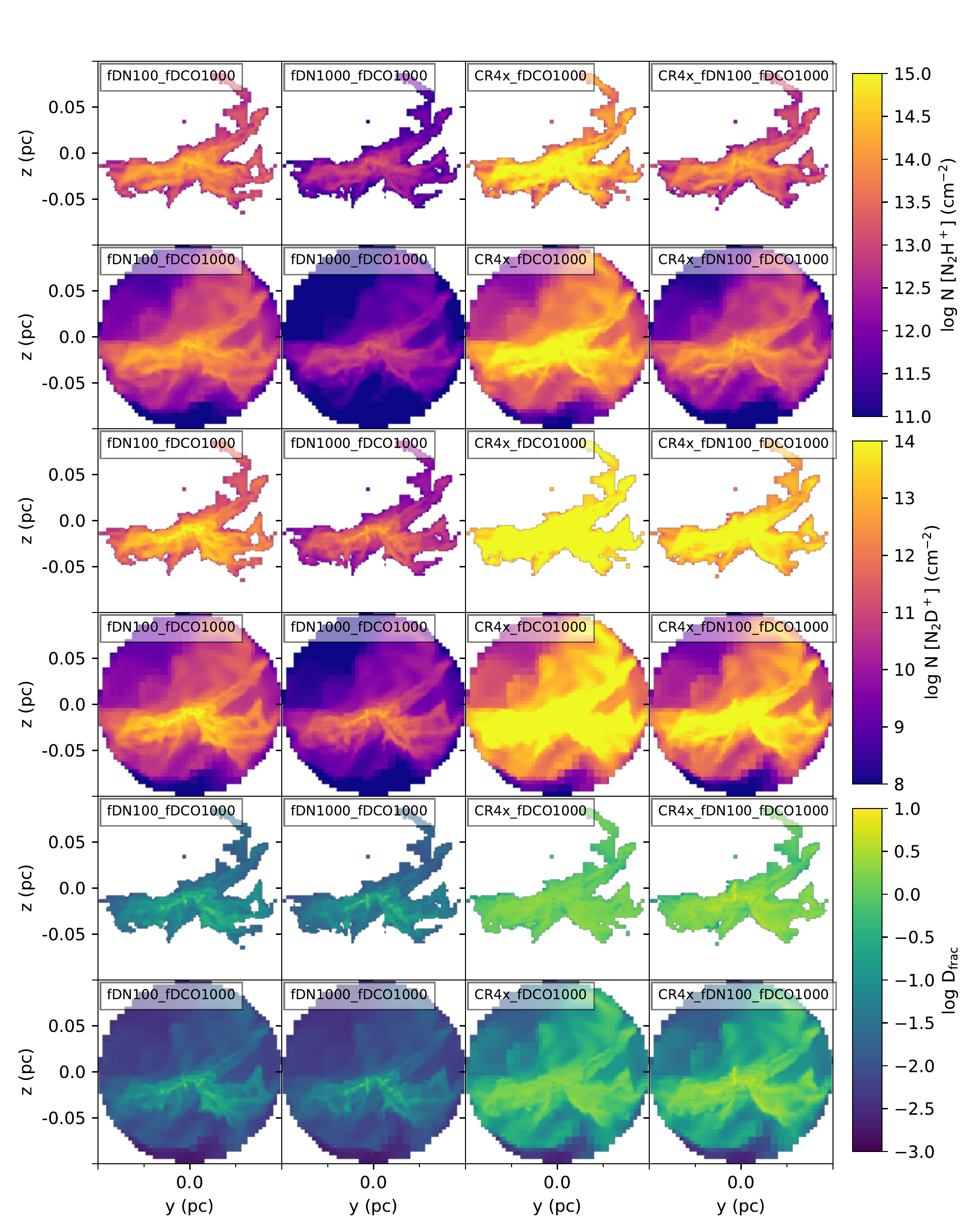}
    \caption{Same as Figure~\ref{fig:dfracmap1-x}, but now for models (left to right): fDN100\_fDCO1000; fDN1000\_fDCO1000; CR4x\_fDCO1000; CR4x\_fDN100\_fDCO1000, with their parameters listed in Table~\ref{tab:chemmodel}. Due to the high levels of {\NNDp} and {\DfracNNH} in these models, different scale bars for {\NNDp} and {\DfracNNH} are used in this figure.}
\label{fig:dfracmap3_x}
\end{figure*}

To investigate the influence of {\N} depletion, we run simulations using two higher {\N} depletion factors ($f_{D}^{\rm N} = 100, 1000$), and with {\CO} depletion factor set to 1000. The results are plotted in the left two columns of Figure~\ref{fig:dfracmap3_x}. It is easy to see that the column densities of {\NNHp} and {\NNDp} are reduced and the spatial structures of {\DfracNNH} are enhanced (note that in this figure the range of {\DfracNNH} now extends up to values of 10).

\subsubsection{Combining higher cosmic-ray ionization rates with other models}

In \S\ref{sssec:crir}, it was shown that the cosmic-ray ionization rate has a strong impact to the level of {\DfracNNH} and its spatial structure. However, we have not yet investigated its influence on the models with higher depletion factors. Thus combinations of a higher cosmic-ray ionization rate of $\zeta = 1.0 \times 10^{-16}$ with the fDCO1000 and fDN100\_fDCO1000 models are carried out. The results are also plotted in Figure~\ref{fig:dfracmap3_x}. These models do not show a strong effect of further enhancement of the spatial structure of {\DfracNNH}. However, due to the effect of shortening chemical time scales, the enhanced cosmic ray models have generally enhanced levels of {\DfracNNH} by about a factor of 10 than in the lower cosmic ray cases. Final values of {\DfracNNH} that are $>1$ can be achieved in some regions.

It is worth noting that high cosmic-ray ionization rates may lead to high rates of desorption of species from grain surfaces such that the highest values we have used for gas phase depletion factors may not be reasonable in this regime. To investigate this, we have run some test models with the astrochemical network of \citet{Walsh2015}, which includes the influences of thermal desorption, photodesorption and grain surface reactions. Cosmic ray especially contributes to the UV field produced by exciting H2 molecules. These results show that in gas with $\rm \nH=10^6\ cm^{-3}$ at $T=15$~K and cosmic ray ionization rate of $\zeta = 1.0 \times 10^{-16}$ the CO depletion factor reaches $\sim 500$ in 100,000 years. However, there are significant uncertainties in the modeling of cosmic-ray induced desorption processes, so we consider our most extreme case with a CO depletion factor of 1000 under such conditions to be a reasonable choice, especially since much of the core material of interest soon achieves densities several times higher than $\rm \nH=10^6\ cm^{-3}$.
%and cosmic-ray attenuation, it is hard to get the real depletion factor for our cores, but the results show that the parameters have a good consistency.

\subsection{Average Core Properties and Comparison to Observed Cores}
\label{sec:obs}

\subsubsection{Average core abundances and {\DfracNNH}}
% \subsubsection{average core quantities?}
\label{sec:aveprop}

We have seen from the results presented so far that there are different combinations of astrochemical conditions that can help achieve a given value of {\DfracNNH} within a single free-fall collapse time of a massive PSC. To break these degeneracies when comparing to observed systems, we will need to use the full information set available, i.e., measurements of the absolute abundances of {\NNHp} and {\NNDp}, as well as their ratio that defines {\DfracNNH}. Additional constraints on {\CO} depletion factor and cosmic-ray ionization rate are also useful, if available.

\begin{figure*}
    \mbox{}\hfill
    \begin{subfigure}[t]{0.48\linewidth}
        \includegraphics[width=\linewidth]{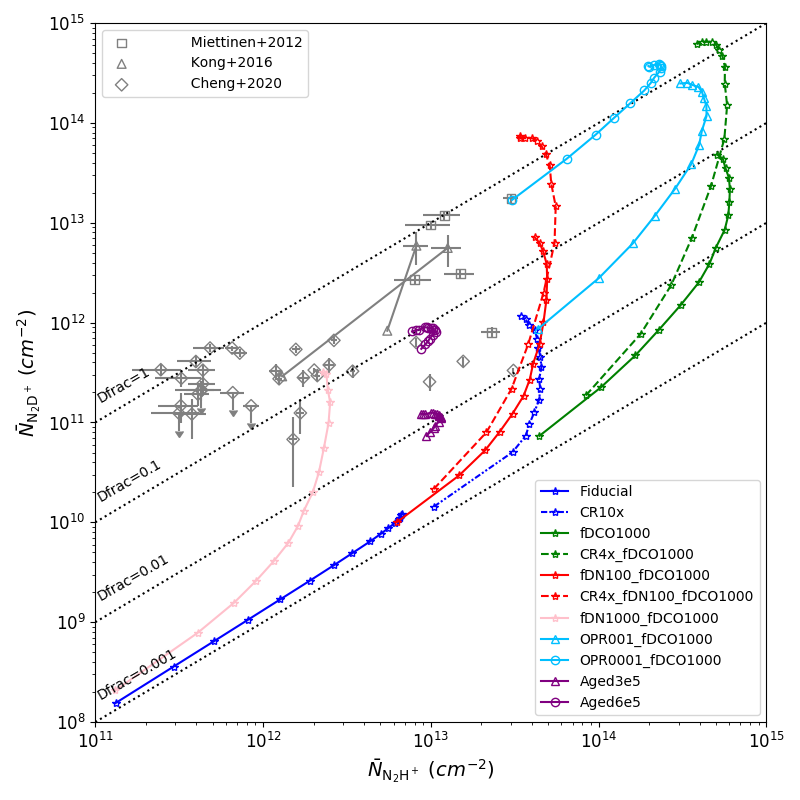}
        % \subcaption{column densities}
        % \label{}
    \end{subfigure}
    \hfill
    \begin{subfigure}[t]{0.48\linewidth}
        \includegraphics[width=\linewidth]{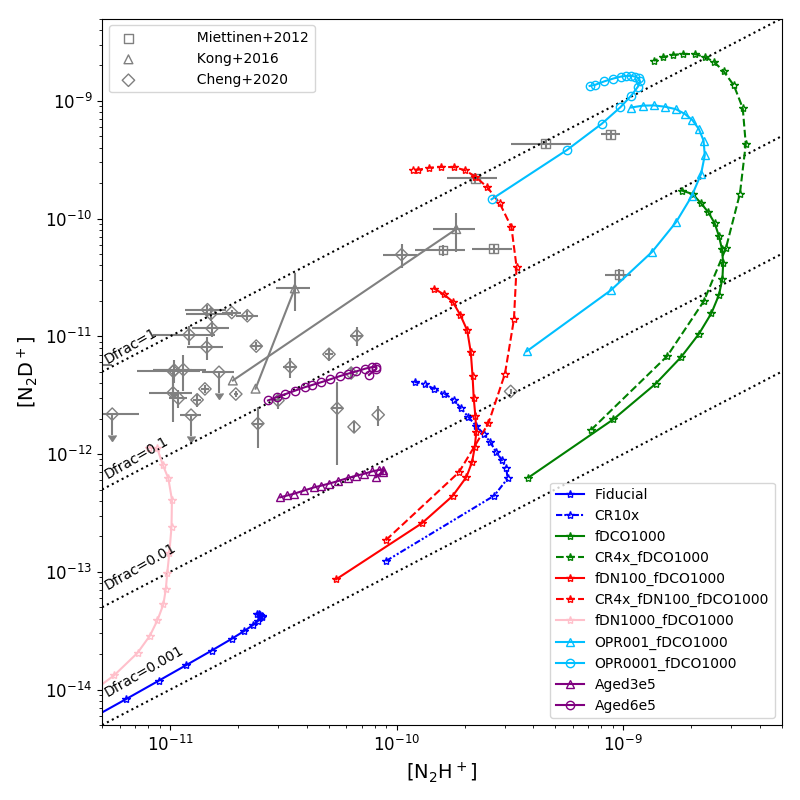}
        % \subcaption{abundances}
        % \label{}
    \end{subfigure}
    \hfill\mbox{}
    \caption{{\it (a) Left:} Time evolution (from 0.05 {\tff} to 0.8 {\tff}) of the average column densities of {\NNHp} and {\NNDp} for various simulations as labelled. The grey squares, triangles and diamonds are observational data from \citet{Miettinen2012, Kong2016} and Cheng et al. (in prep), respectively.
    {\it (b) Right:} As (a), but now showing average abundances of {\NNHp} and {\NNDp} with respect to H nuclei.}
    \label{fig:observation}
\end{figure*}

We calculate the average column densities of {\NNHp} and {\NNDp} in our simulations. Assuming the {\NNDp}(3-2) emission area can be well resolved, we calculate the average over this resolved area of the map that contains {\NNHp} and {\NNDp} emission (recall that these are sight lines that contain at least one cell above the adopted density threshold $\nHcrit = 8 \times 10^5 {\rm cm^{-3}}$). The average column densities of {\NNHp} and {\NNDp} in the core are shown in Figure~\ref{fig:observation}a, with the lines tracking the evolution of the simulations in time, i.e., rising up from the lower left to the upper right. The behaviour of {\NNHp} is that it tends to grow to reach a maximum value and then declines by a small amount. On the other hand, {\NNDp} tends to keep increasing during the evolutionary phases that we follow. The largest factor that leads to variation among the models is the adopted N depletion factor. 

Several observed cores and sub-regions of cores are also shown in Figure~\ref{fig:observation}a. These exhibit values of column densities and {\DfracNNH} that are similar to those of some of the simulations. We note that many of the observed cores are of relatively low-mass (e.g., those of Cheng et al., in prep. and many of the sources of \citet{Miettinen2012}) and low total H column, so that one does not necessarily expect agreement in terms of absolute columns of {\NNHp} and {\NNDp} with the simulations, which are for a relatively high density, 60~$M_\odot$ core. Thus, we shall also focus on abundances of the species relative to H.

To estimate the H column density in the simulations to be used in the abundance estimates, which are shown in Figure~\ref{fig:observation}b, we only count material from cells above the {\nHcrit} density threshold, with an additional constraint that the material be located within 0.1~pc distance of the core centre, i.e., excluding surrounding clump environment material. We note that when such measurements are done observationally, then, to be similar to this theoretical method, the clump background contribution should be subtracted off the estimate of the core region. As expected, the simulations show qualitatively similar behaviour in their abundance evolution as shown in their absolute column density evolution. 
%We also see that the observed cores span a range of abundances that are more closely matched to the ranges explored in the simulations, where N depletion factor is again one of the main variables setting the ranges found in the models.

%In Figure~\ref{fig:observation}, the density threshold ($\nHcrit = 8 \times 10^5 {\rm cm^{-3}}$) is applied to {\NNHp, \NNDp} and {\Hy}. 
%For the average column densities of {\NNHp} and {\NNDp} plotted in the left panel of Figure~\ref{fig:observation}. 

We note that if a simpler circular aperture with a 0.1~pc radius
% cjhsu - should be 0.1~pc leading to the decrease of column density
% same effective area 
is used in the analysis, then the estimates of absolute column densities of {\NNHp} and {\NNDp}  generally decrease, typically by factors of 2 to 3, since significant area from non-emitting regions is included. However, the estimates of the abundances with respect to H change by smaller amounts.
%Since we may not always guarantee sufficient resolution in observation, the curves could shift to left when considering a finite size resolution. For example, if the aperture has a size of 0.1~pc circle, the column densities will reduce by around a factor of 2.8.
The density threshold is another factor influencing the average column densities. If we remove the density threshold but still limit the valid domain inside the 0.1~pc radius sphere, the column densities of {\NNHp} can grow by factors of $1.3$ to $3.2$ (time-averaged in the various models). Similarly, the column densities of {\NNDp} grow by factors of $1.1$ to $3.2$, but always by a slightly smaller factor than {\NNHp}. For the column density of H, we find that $N_{\Hy}$ increases by a factor of 1.74 when the density threshold condition is relaxed.

Overall, considering the comparison of simulations with observed cores, we see that the observed cores exhibit a range of values of {\meanNNNHp} from about $10^{11}$ to a few $\times 10^{13}\:{\rm cm}^{-2}$ and of {\DfracNNH} from a few $\times 0.01$ to about 1. It is to be remarked that there can be significant uncertainty in the estimates of the observed column densities, e.g., due to excitation temperature uncertainties \citep[e.g., see the case of the C1-S and N cores observed by][]{Kong2016} or the cores of \citep[][]{Miettinen2012}, where a mixture of LTE and non-LTE assumptions were employed. Estimates of the abundances of the species bring in additional uncertainties associated with the measurement of $N_{\rm H}$. The observed values of [$\rm {N_2H^+}$] range from about $10^{-11}$ to $10^{-9}$, while those of [$\rm {N_2D^+}$] range from about $10^{-12}$ to close to $10^{-9}$. Focusing on the abundance plots, we see that the range of observed core properties is well covered by the simulations. For example, the model CR4x\_fDN100\_fDCO1000 reaches the high values of {\DfracNNH} that are observed in some cores with typical abundances of $\rm [N_2H^+]\sim {\rm few} \times 10^{-10}$. We see that variation in $f_D^N$, i.e., the parameter controlling the degree of {\N} depletion from the gas phase, is one way that the simulations can be tuned to match a given observed system. Given the uncertainties in predicting the degree of N depletion from first principles, one strategy when modeling a population of cores in the same region (assumed to have similar environmental conditions) is to set this empirically based on observations of {\NNHp} and then use any observed differences in {\NNDp} abundance as a measure of evolutionary stage, with absolute age differences estimated from such simulations.

%For the observed systems with very high values of {\DfracNNH}, i.e., close to unity, fDN100\_fDCO1000 and CR4x\_fDN100\_fDCO1000 models provide a possible way to match with those values.

%The results could be checked if we can observe these objects with higher resolution.
% show a typical range of tracer column density between $10^{12}$ to $10^{14}\ \mathrm{cm^{-2}}$ \citep{Kortgen2017}. For the H\_DeCO model, either one of the tracers shows a much higher column density than the observational data. H\_CR model could be more possible to exist in real world. \par

\subsubsection{Spatial structure of {\DfracNNH}}\label{sec:spatial}

We have seen that different models lead to different degrees of spatial concentration of {\DfracNNH} that follow the PSC structure, i.e., the level of deuteration can increase rapidly with density so that the PSC appears as an "island" of high {\DfracNNH} within the surrounding clump environment. While there are to date relatively few observed maps of {\DfracNNH}, especially on the scale of PSCs, in a number of observational studies, the morphology of {\NNDp} emission appears more concentrated than that of {\NNHp} \citep[e.g.,][]{Kong2016,Barnes2016}. Among the models that we have considered so far, those having high \CO\ depletion factors or high cosmic-ray ionization rates show significant degrees of spatial concentration. In the last sub-section, we compared core averaged column densities and abundances with observational data and found that the CR4x\_fDN100\_fDCO1000 model is one of the best of those explored so far in matching the observed systems.  
%In our section 3, we mainly focus on the results of {\DfracNNH}. Besides checking the values of {\DfracNNH}, we should also care about the absolute abundances of {\NNHp} and {\NNDp}. 
%Because CR10x and fDCO1000 models show high deuterium fractions with clearer spatial structures in our simulations. We would like to check the absolute abundances of these two cases. 

\begin{figure*}

%   \begin{subfigure}{\linewidth}
%     \centering
%     \includegraphics[scale=0.35]{fig/scatterpanel3_nh_x_N2HN2D_t8_DN100.png}
%     \caption{fDN100\_fDCO1000 model}
%     \label{absabCR10}
%   \end{subfigure}

%   \begin{subfigure}{\linewidth}
%     \centering
%     \includegraphics[scale=0.35]{fig/scatterpanel3_nh_x_N2HN2D_t8_CR4_DN100.png}
%     \caption{CR4x\_fDN100\_fDCO1000 model}
%     \label{absabDCO1000}
%   \end{subfigure}

\centering
\includegraphics[scale=0.35]{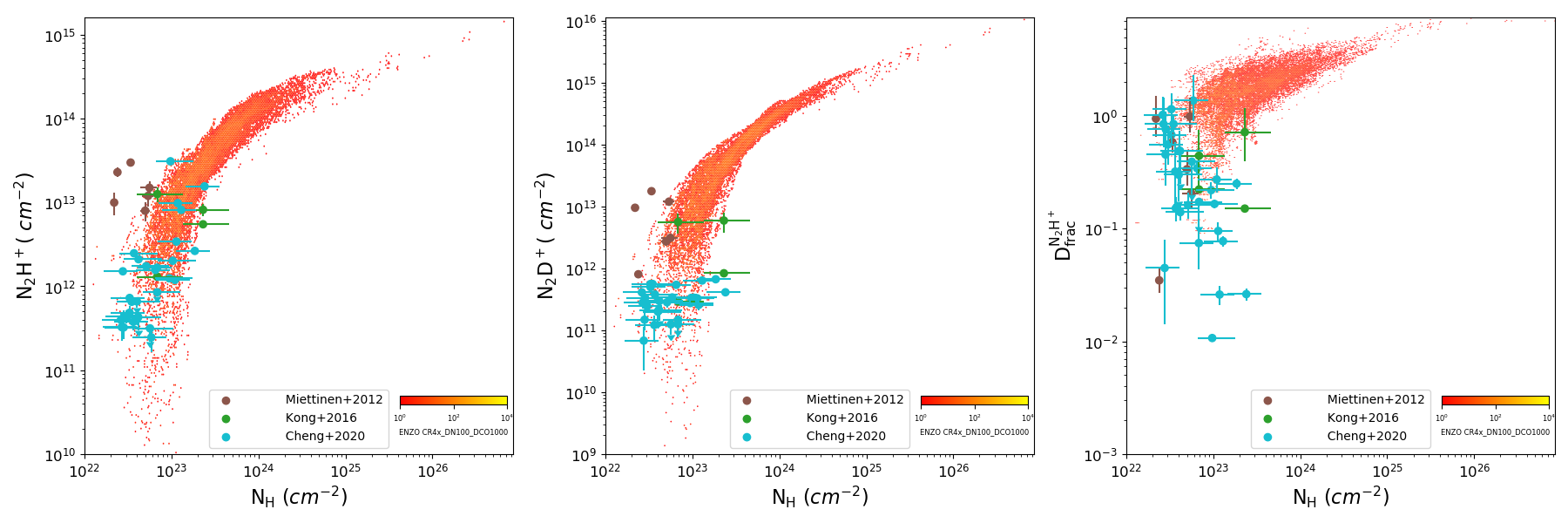}
\includegraphics[scale=0.35]{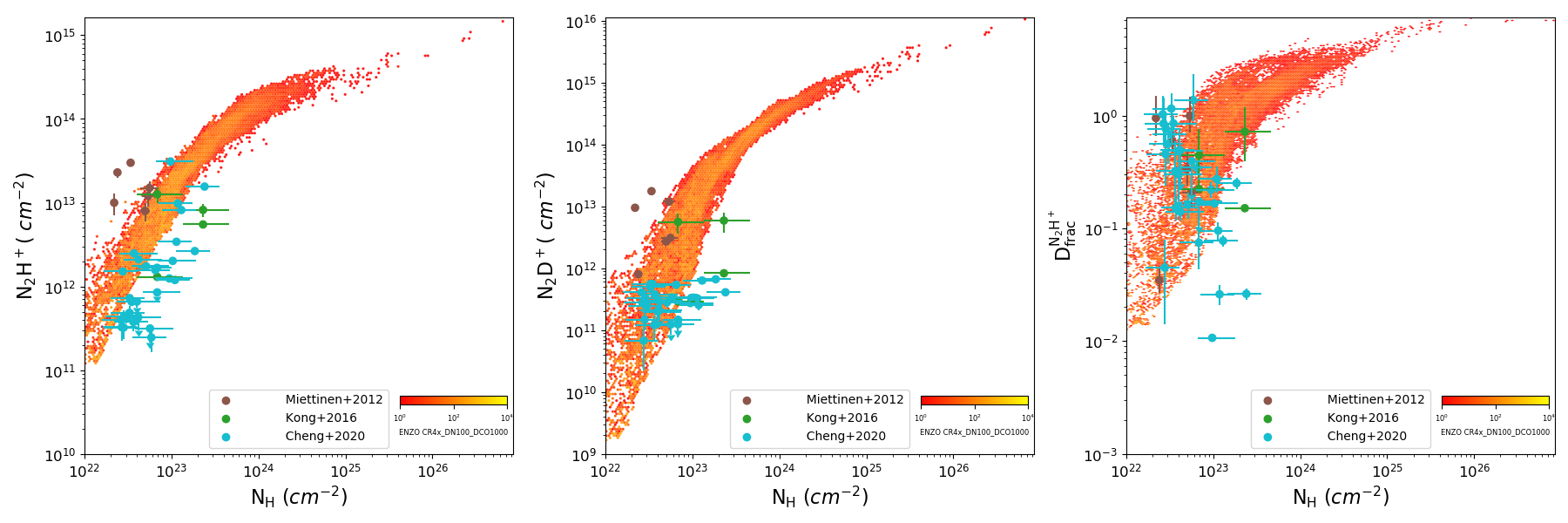}
\caption{The panels from left to right show the column density of {\NNHp}, the column density of {\NNDp} and {\DfracNNH} with respect to the column density of H nuclei ($N_{\rm H}$) in the CR4x\_fDN100\_fDCO1000 model. The top row shows results based on the method of requiring a density threshold for {\NNHp} and {\NNDp} emission, while the bottom row shows the case without this threshold. Observational data points, which are the same as presented in Figure~\ref{fig:observation}, are plotted as labelled.}
\label{fig:absab}
\end{figure*}

In Figure~\ref{fig:absab} we examine the dependence of {\NNHp} and {\NNDp} column densities and {\DfracNNH} on total H column density, $N_{\rm H}$, in the %fDN100\_fDCO1000 and 
CR4x\_fDN100\_fDCO1000 model at $t=0.8 t_{\rm ff}$ for two cases of with and without the density threshold being needed for {\NNHp} and {\NNDp} emission. We see that there is a general rise in these quantities with $N_{\rm H}$. The use of a density threshold in the simulations for {\NNHp} and {\NNDp} emission causes a truncation at low values of $N_{\rm H}$, as expected.

%The ranges of the column density of {\NNHp} that are covered are from $10^{11}$ to $10^{15} {\rm (cm^{-2})}$ except for some extreme values above $10^{15} {\rm (cm^{-2})}$. In the contrast, the column density of {\NNDp} ranges between $10^{9}$ and $10^{15} {\rm (cm^{-2})}$ in fDN100\_fDCO1000 model but mostly ranges between $10^{10}$ to $3\times10^{15} {\rm (cm^{-2})}$ in CR4x\_fDN100\_fDCO1000 model. The difference can also be observed from the {\DfracNNH} panels. CR4x\_fDN100\_fDCO1000 model has the minimum values 10 times higher than those in fDN100\_fDCO1000 model and has a higher tail. 

Figure~\ref{fig:absab} also shows observational data from various cores, i.e., the same data shown in Figure~\ref{fig:observation}. These data include core-averaged properties, as well as a small strip map from Cheng et al. (in prep.), which probes a range of conditions in a linear feature in the Vela C GMC that spans from two early-stage protostellar cores and a bridge feature containing a PSC candidate in between. These systems are not as massive as our simulated core and are also generally of lower column density, as can be seen in Figure~\ref{fig:absab}. Nevertheless, they provide a number of probes of independent regions that span conditions relevant to the outer boundary of our simulated massive PSC. The comparisons that are shown in Figure~\ref{fig:absab} indicate a preference for not using the density threshold condition for {\NNHp} and {\NNDp} emission. Still, a self-consistent test of the model will require new observations to make resolved maps of {\NNHp} and {\NNDp} of a massive PSC.

% \begin{figure*}
% \centering
% \includegraphics[scale=0.35]{fig/scatterpanel3_nh_x_N2HN2D_t8_DCO1000.png}
% \caption{The absolute abundances of {\NNHp} and {\NNDp} of H\_DeCO model at 0.8 $t_{ff}$.}
% \label{absabDCO1000}
% \end{figure*}

% \begin{figure*}
% \centering
% \subfigure{
%     \includegraphics[scale=0.5]{fig/scatter2d_nh_x_N2H_t8_DCO1000.png}
% }
% \subfigure{
%     \includegraphics[scale=0.5]{fig/scatter2d_nh_x_N2D_t8_DCO1000.png}
% }
% \caption{The absolute abundances of {\NNHp} and {\NNDp} of H\_DeCO model at 0.8 $t_{ff}$.}
% \label{absabDCO1000}
% \end{figure*}

\subsubsection{Chemical history of the core defined by {\NNDp}}

In \S\ref{sec:aveprop} and \S\ref{sec:spatial} we have considered the time evolution of average column densities, abundances and the spatial structures of different simulated cores. We then selected the CR4x\_fDN100\_fDCO1000 model as one of the best cases for matching some highly deuterated observed systems. We now focus on the time evolution of the average properties of this core.

\begin{figure}
    \centering
    \includegraphics[width=\linewidth]{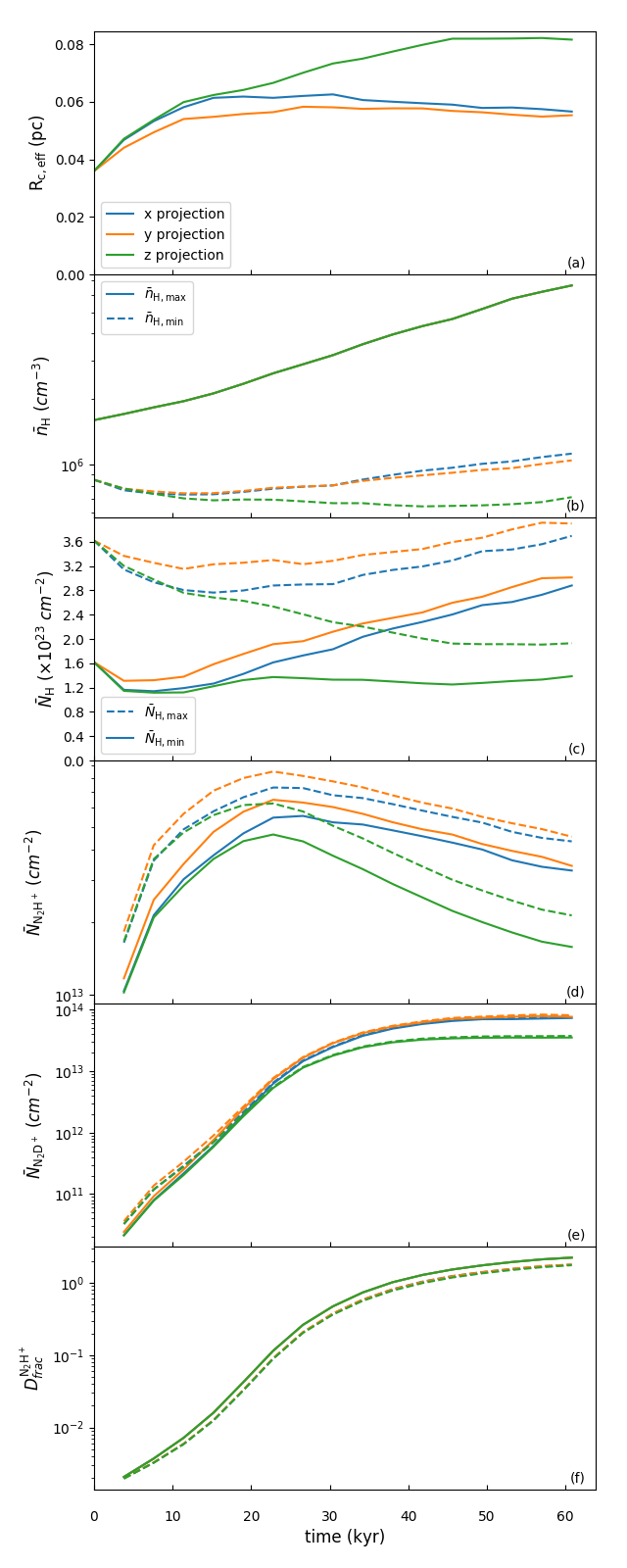}
    \caption{Time evolution of average quantities based on the {\NNDp} emission area (see text) of CR4x\_fDN100\_fDCO1000 model: (a) effective radius (here and in subsequent panels, results for $x$, $y$, $z$ projection are shown with colors as labelled); (b) average number density of H nuclei ($\bar{n}_{\rm H}$) (here and in subsequent panels solid line(s) show results considering only the cells where $\nH>\nHcrit$; dashed lines are results when other, lower density cells along the line of sight are included in the average); (c) average column density ($\bar{N}_{\rm H}$); (d) average column density of {\NNHp}; (e) average column density of {\NNDp}; (f) average {\DfracNNH}.}
    \label{fig:abtseries}
\end{figure}

In a number of observational studies \citep[e.g.,][]{Tan2013,Kong2018}, the PSC boundary is defined by the region (either in position-position space or in position-position-velocity space) that is detected in {\NNDp}(3-2) line emission. In our simulations, following \citet{Goodson2016}, we have assumed {\NNDp} emission arises only if the H nuclei number density of a cell is larger than $8\times10^5\:{\rm cm}^{-3}$). We thus measure the effective area of the core, $A_{\rm c,eff}$, in a given image as the area of all the pixels that contain at least one cell along the line of sight that meets this density threshold. Then, we define the effective radius of the core as $R_{\rm c,eff}\equiv (A_{\rm c,eff}/\pi)^{1/2}$. The time evolution of the effective radii as viewed from different directions are plotted in Figure~\ref{fig:abtseries}a. 
% If the line of sight is along the $x$ or $y$ axis, the radius becomes stable at around 0.05~pc after 0.3 {\tff}. In contrast, the radius becomes stable after 0.6 {\tff} when the core is viewed from $z$ axis.
In the case of the $x$ and $y$ projections, we see that the core radius grows quickly from an initial value of 0.037~pc to about 0.055~pc, being fairly constant (or even declining) after $t=0.2 t_{\rm ff}$. In the case of the $z$ projection, $R_{\rm c,eff}$ grows from 0.037~pc to about 0.08~pc, which is reached by $t=0.6 t_{\rm ff}$, after which it remains nearly constant. The larger effective area when viewed in the $z$-projection (i.e., viewing the $x-y$ plane) is clear from Figures~\ref{fiducial_x} and \ref{fiducial_z}.

Figures~\ref{fig:abtseries}b and c show the time evolution of the average H nuclei number density, {\meannH}, and the average total H column density, {\meanNH}, for the core defined by its {\NNDp} emitting region. Two methods for estimating densities and column densities are used. The first method, which was introduced in \S\ref{sec:aveprop}, only counts cells that satisfy the density threshold condition and that are located within a 0.1~pc radius of the center of the simulation domain. This criterion counts only the densest and the most limited part inside the core resulting in the $\bar{n}_{\rm H,max}$ and $\bar{N}_{\rm H,min}$. However, since from an observational point of view it is hard to make such a selection based on such a density threshold, in our second method this density threshold condition is removed, while still keeping the 0.1~pc radius constraint. This estimate gives us $\bar{n}_{\rm H,min}$ and $\bar{N}_{\rm H,max}$ because of the extended low density region. It is noticeable that the average volume density estimated by the latter method remains steady at around $10^6 {\rm (cm^{-3})}$ during the course of the simulation. The average density in the actual {\NNDp}-emitting cells is about twice as large at the beginning and seven times larger at the end. 
%The growth of the average volume density in the densest cells indicates the collapse efficiency of the core.
Similarly, the two methods give a range of column densities, with the differences being at the level of about a factor of three at early times and less than a factor of two at late times. Overall, $\bar{N}_{\rm H}$ stays in a range from $10^{23}$ to $4\times 10^{23}$. For the first method, i.e., with the density threshold condition (solid lines), $\bar{N}_{\rm H,min}$ keeps growing in the $x$ and $y$ projections, but is almost constant in the $z$ projection. For the second method, i.e., without the density threshold condition (dashed lines), $\bar{N}_{\rm H,max}$ is relatively constant (but with a slow, later time increase) in the $x$ and $y$ projections, while in the $z$ projection it steadily decreases during most of the evolution of the simulation.

% In Figure~\ref{fig:abtseries}c, the time evolution of the average total H column density, $N_{\rm H}$, estimated by two methods are plotted. Both methods estimate column densities inside the {\NNDp} emitting area. The first method, which was introduced in \S\ref{sec:aveprop}, is a direct sum of the H nuclei in each image pixel that defines $A_{\rm c,eff}$, but only counting the cells that satisfy the density threshold condition and that are located within a 0.1~pc radius of the center of the simulation domain. We refer to this estimate as $N_{\rm H,min}$. However, since from an observational point of view it is hard to make such a selection based on such a density threshold, in our second method this density threshold condition is removed, while still keeping the 0.1~pc radius constraint. We refer to this estimate as $N_{\rm H,max}$. The two methods give us a range of column densities, with the differences being at the level of about a factor of three at early times and less than a factor of two at late times. Overall, $N_{\rm H}$ stays in a range from $10^{23}$ to $4\times 10^{23}$. For the first method (solid lines), $N_{\rm H,min}$ keeps growing in the $x$ and $y$ projections, but is almost constant in the $z$ projection. For the second method (dashed lines), $N_{\rm H,max}$ is relatively constant (but with a slow, later time increase) in the $x$ and $y$ projections, while in the $z$ projection it steadily decreases during most of the evolution of the simulation.

In Figures~\ref{fig:abtseries}d, e and f, the column densities of {\NNHp} and {\NNDp} and the implied value of {\DfracNNH} are examined during the evolution of the core and with the two methods of total H column density estimation, i.e., with and without the use of the density threshold. The column densities of {\NNHp} are seen to quickly establish relatively high values and then undergo a modest decline, while those of {\NNDp} undergo a more gradual rise and then reach near constant levels. 
%Differences in the methods of column density estimation make relatively smaller differences to these results given than the {\NNHp} and {\NNDp} species are concentrated in the higher density regions. 
There are relatively modest variations due to different methods of column density estimation and different viewing angles.

One promising direction to improve observational constraints is to survey a sample of PSCs in the same region, i.e., assumed to have similar cosmic-ray environmental conditions, and examine the distribution of {\NNHp} column densities to determine the maximum achieved value and the distribution up to this value. This information can then potentially better constrain the types of acceptable astrochemical models (including a combination of cosmic-ray ionization rates and achieved levels of N species depletion) that are valid for the cores in the region. Then associated measurements of {\DfracNNH} can further constrain the astrochemical conditions, along with the dynamical evolution of the cores. In the next sub-section we examine the detailed kinematic and dynamical properties that can be derived for the core based on observations of its {\NNDp} emission.

%From the three directions, it is noticeable that the species column densities change more apparently in $z$ direction but very close in $x$ and $y$ direction. This matches with the fact that the magnetic field breaks the symmetry in the $z$ direction and makes the effective area larger. However, they show the same {\DfracNNH} from any angle since the same cells are considered. Overall, {\NNHp} stays in a range from $10^{13}$ to $10^{14}$. The peak value $\sim 5\times10^{13}$ appear at around 0.3 {\tff}. In contrast, {\NNDp} grows several scales (from $10^{12}$ to $10^{14}$) during 0.05 {\tff} to 0.5 {\tff}.

\subsection{Core Structure, Kinematics and Dynamics}\label{S:dynamics}

Our goal here is to consider the structure, kinematics and dynamics of the PSC, especially as traced by its {\NNDp} emission. Note, all the simulations we have run have effectively the same dynamical properties, and indeed start with the same physical conditions of the particular choice of turbulent velocity field. However, of the models that we have presented so far, we consider that the CR4x\_fDN100\_fDCO1000 is best at reproducing observed features of massive PSCs. In particular, it can reach high levels of {\DfracNNH} $\gtrsim 1$ and also do this in a way that {\DfracNNH} increases with density (or column density), so that the core appears as a local maximum in the {\DfracNNH} map. However, it remains to be established if its kinematic properties, as traced by {\NNHp} and {\NNDp}, match those of observed systems. Thus, here we focus on the kinematic and dynamical properties of the core in the CR4x\_fDN100\_fDCO1000 model as traced by these species.

\subsubsection{Evolution of Core Size and Associated Mass}

The evolution of $R_{\rm c,eff}$ versus time in the simulation for the CR4x\_fDN100\_fDCO1000 model has been shown in Figure~\ref{fig:abtseries}a, including for the three different viewing directions.
To estimate the mass associated with the core, we apply the two methods, i.e., with and without density threshold, that we used to estimate column densities in \S\ref{sec:obs}. Note, both methods require the core material to be within 0.1~pc distance of the initial center of core domain.

%consider the emission in the 0.1~pc radius sphere. The difference between the two methods is the first one is limited by the density threshold, but the second is not. The two methods give us the lower and upper limit estimation of the core mass.

\begin{figure*}
    \centering
    \includegraphics[width=\linewidth]{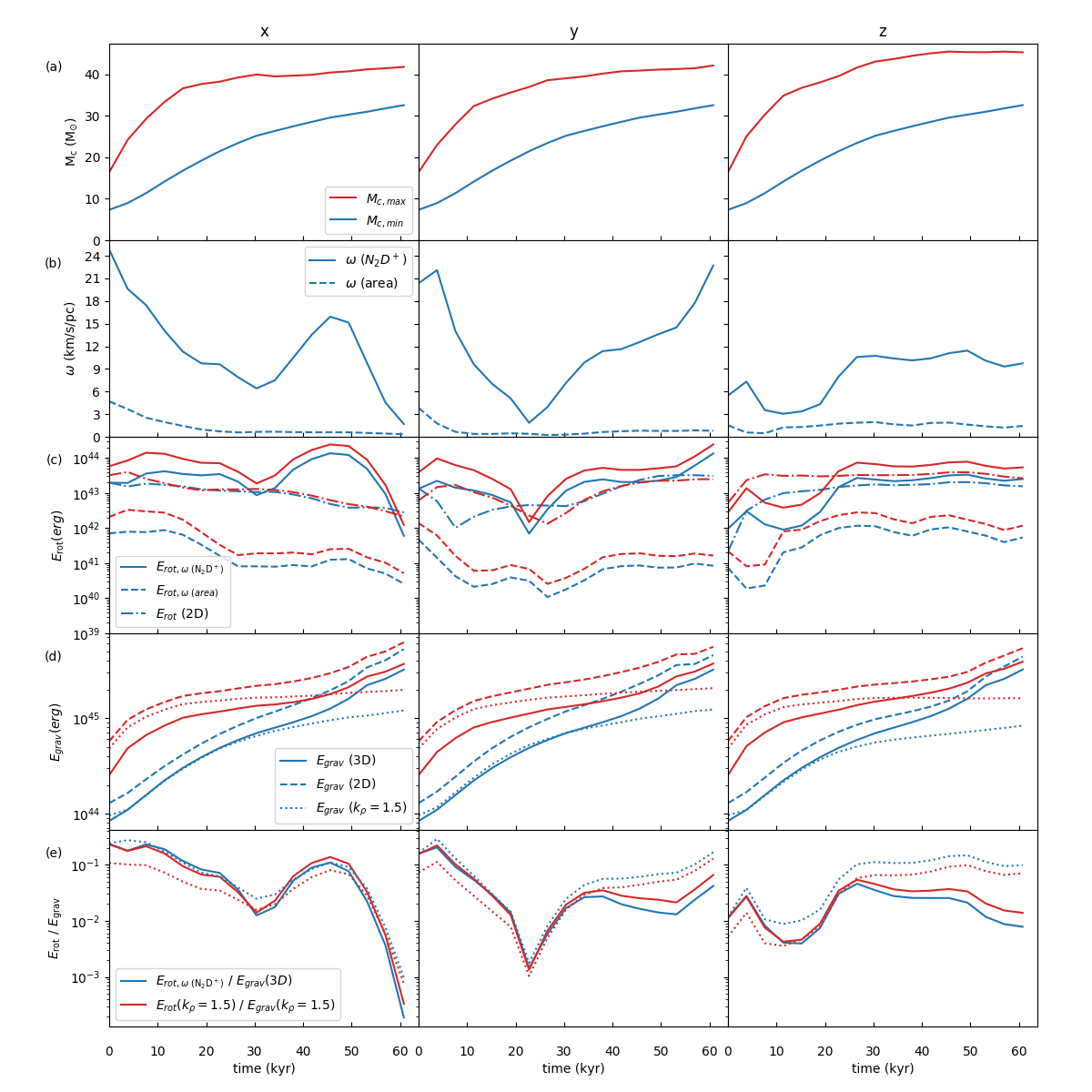}
    \caption{Time evolution of PSC kinematic and dynamic properties. The results based on different viewing directions, i.e., projecting along the $x$, $y$ and $z$ axes, are shown in the left, middle and right columns, respectively. From top to bottom: (a) core masses, $M_{\rm c,min}$ (blue line), estimated using only cells that meet the density threshold condition, and $M_{\rm c,max}$ (red line), estimated including lower density material along the line of sight (see text); (b) velocity gradient, $\omega$, of {\NNDp}-weighted radial velocities (solid line) and area-weighted radial velocities (dashed line); (c) rotational energy, $E_{\rm rot}$, estimated from (i) {\NNDp}-weighted velocity gradient and projected 2D plane moment of inertia (solid lines), (ii) area-weighted velocity gradient and projected 2D plane moment of inertia (dashed lines), (iii) the total angular momentum of each projected plane (dot-dash line); 
    %(iv) the angular momentum contribution of radial velocity in 3D (dotted line); 
    (d) gravitational energy estimated from (i) a simple polytropic model (dotted lines, $k_\rho = 1.5$), (ii) pairwise summation of cells in the 2D plane (dashed lines), (iii) pairwise summation of cells in 3D (solid lines); and (e) ratio of rotational energy to gravitational energy estimated from (i) 2D plane method (solid lines), (ii) polytropic model method (dotted lines).
    }
    \label{fig:rottseries}
\end{figure*}

The time evolution of the core mass, $M_c$, measured in these different ways, including using different viewing directions, is shown in Figure~\ref{fig:rottseries}a. From the physical model of a $60\:M_\odot$ core, we see that the mass potentially traced by {\NNDp} increases from about $8\:M_\odot$ at the beginning of the simulation to reach just over $30\:M_\odot$ in the case of the first method requiring a density threshold (note, the same mass is derived independent of viewing direction). With the second method, the traced core grows from $17\:M_\odot$ to reach just over $40\:M_\odot$ by the end of the simulation (with minor variations caused by different viewing directions).

\subsubsection{First Moment Maps, Velocity Gradients and Implied Rotation}
\label{sec:firstmom}

\begin{figure*}
    % \centering
    \includegraphics[width=\linewidth]{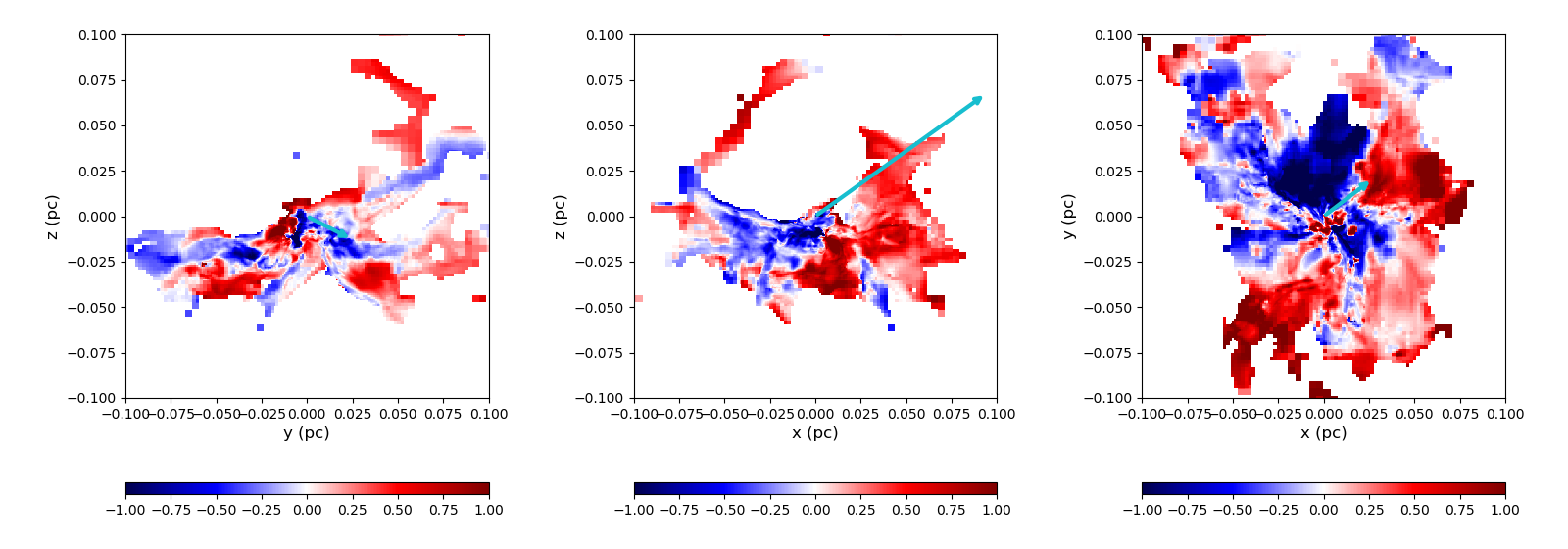}
    \caption{The first moment map along each principal axis at t=0.8$t_{\rm ff}$. The cyan vectors indicate the directions and magnitudes of the average velocity gradients weighted by {\NNDp} flux on each plane.}
\label{fig:mom1}
\end{figure*}

In Figure~\ref{fig:mom1}, we show the first moment maps as viewed along each principal axis, weighted by {\NNDp}, of the core from simulation CR4x\_fDN100\_fDCO1000 at the final time corresponding to $t=0.8t_{\rm ff}$. Note that the large scale $B$-field is initially in the direction of the $z$-axis and we have already seen that the collapse occurs most easily along the direction of this axis. Thus the spatial distribution of the dense gas and the {\NNDp} emission is more concentrated towards the $z=0$ plane.

We use the first moment maps to evaluate the implied average velocity gradient (magnitude and direction) in each image plane, with the averaging done using the column densities of {\NNDp}. Such a gradient can be interpreted as rotation of the core with an angular momentum, $J$, vector in the plane of the sky. These vectors are also shown in each of the panels.
%The bottom two rows show the position-mean velocity relation along each direction. Each curve is fitted by a straight line to estimate the velocity gradient across the core. The slopes of the fitted lines give us the velocity gradient in each direction and thus the overall velocity gradient and its orientation in the plane of the sky can be evaluated. The total velocity gradients we derive from the three maps are plotted on the maps by the green vectors. 
The magnitudes of the velocity gradients are 5.57, 23.1, and 6.73~km~$\rm s^{-1}\:pc^{-1}$ for the $y-z$, $x-z$ and $x-y$ planes, respectively. If we define north to be 0$^{\circ}$ in each panel, i.e., along the $z-$axis in the first two panels and along the $y-$axis in the third panel, then their position angles are -117$^{\circ}$, -54$^{\circ}$, and -53$^{\circ}$, respectively. Since {\NNDp} is concentrated in the central region, the velocity gradients are dominated by the small area in the center. Overall, the kinematics traced by the first moment maps of {\NNDp} appear relatively disordered, which likely reflects the fact that the core is undergoing approximately free-fall global collapse from turbulent initial conditions, i.e., the collapse is only modestly impeded by the support of turbulence and magnetic fields.

In Figure~\ref{fig:rottseries}b, we show the time evolution of the magnitudes of the velocity gradients of the core. Two ways are used to estimated the velocity gradient. The first way, illustrated in Figure~\ref{fig:mom1}, estimates the velocity gradient from the {\NNDp}-weighted radial velocities. For a second method, we use an area-weighted radial velocity from the first moment map to estimate a gradient. In Figure~\ref{fig:rottseries}b, it is clear that the area-weighted velocity gradient is much smaller than the {\NNDp}-weighted velocity gradient by about one order of magnitude. Since the strongest gradients and the highest densities are concentrated near the center, the overall average core velocity gradient is flattened if it is weighted by area. 

There are relatively few spatially resolved massive pre-stellar and/or early-stage protostellar cores observed in {\NNDp}. For the early-stage protostellar core C1-Sa, \citet{Kong2018} estimated a velocity difference of 0.8~km/s across its 0.044~pc diameter, i.e., a velocity gradient of about $18\:{\rm km\:s^{-1}\:pc^{-1}}$. Such a gradient should be compared to our area-weighted estimates: the velocity gradient of C1-Sa is much larger than we see in our PSC simulation. The reason for this could be that the gradient in C1-Sa is already dominated by the presence of the protostar and/or that it is an overall more compact structure (by a factor of about 3) compared to our simulated PSC. On the other hand, for the PSC C1-S, which at its outer scale has a radius of 0.045~pc, \citet{Kong2018} did not discern any significant velocity gradient in the {\NNDp} first moment map. From their Figures 9j and 9n, we see that most of the radial velocities are within 0.1~km/s of the core centroid velocity, but with a few localised regions, especially near the outer boundaries, that show velocity excursions up to about 0.5 km/s. Overall, a qualitiative visual comparison between the first moment maps of our simulated core and those of C1-S suggests that C1-S is more quiescent, which likely indicates that it has a  lower level of internal turbulence and/or smaller infall motions.

In the survey of \citet{Kong2017}, the {\NNDp}-defined core C9A shows a strong velocity gradient in {\NNDp}(3-2): on one side the mean velocities are -1.5~km/s, whereas they are +0.5 km/s on the other. The core diameter is about 0.1~pc, so the velocity gradient is $\sim20\:{\rm km\:s^{-1}\:pc^{-1}}$. \citet{Kong2017} speculated that given the bimodal morphology of its 0th-moment map, it is possible that two {\NNDp} cores are being seen in the process of merging. However, the magnitude of this velocity gradient is within the range seen for our simulated single PSC. The other large cores in the \citet{Kong2017} sample, i.e., B1A, B1B and H2A, do not show such large velocity gradients as C9A.

The average velocity gradients can be interpreted in terms of global rotation of the core and thus used to estimate the rotational energy: $E_{\rm rot} = (1/2) I \omega^2 = J^2/(2I)$, where $I$ is the moment of inertia.
%we have to find the momentum of inertia corresponding to the measured angular velocity. 
%Observationally, due to the limitation of resolution, the momentum of inertia could be estimated by uniform sphere, which gives formula
In many observational cases, especially those with limited resolution, a simplifying assumption for density distribution is often needed for the calculation of $I$. For example, for a uniform sphere $I=(2/5)M_c R_c^2$, while for a singular polytropic sphere $I = (2/3)[(3-k_\rho)/ (5-k_\rho)] M_c R_c^2$ for a given power law index $k_\rho$. 

For our simulated core, we can obtain the actual moment of inertia from the full density distribution. However, we choose to work with the 2D mass surface density distribution, which in principle could be observed in real PSCs, e.g., via high resolution extinction mapping, rather than with the 3D density distribution. Also the average velocity gradients that are derived from line-of-sight radial velocities yield a core rotational axis that is restricted to be in the plane of the sky. Thus we evaluate $I$ with reference to this rotational axis, which runs through the centre of mass of the core as viewed in a given mass surface density map, i.e., for a given viewing direction. 
%From the projected plane, since we have different velocity gradients estimated from the two methods, the velocity gradients then are defined as the rotational axis on each plane. 
Thus, the defined core area is treated as a thin slab rotating along the axis.

With this method for estimating $I$, the rotational energy evolution of the core is shown in Figure~\ref{fig:rottseries}c for the four combinations of two mass cases (red and blue colors) and two velocity gradient cases (solid and dashed line styles), described above. One sees that the different methods of estimating velocity gradients translate into about two orders of magnitude differences in the amount of rotational energy.

Next, we compare these estimates to the actual rotational energy of the core, but with this estimate still based on radial velocities along the line of sight to the projected plane, i.e., 
%The moment of inertia can be found by the distribution of column density. 
we sum the angular momenta of all the 3D cells with respect to the projected center of mass. Then the total rotational energy is estimated via $J_{\rm tot}^2/(2I)$. This estimate is shown with the dot-dashed lines in Figure~\ref{fig:rottseries}c. 
%A variation on this method uses the 3D distance of each cell to the rotation axis, which results in a slightly larger value of $I$ and thus lower value of $E_{\rm rot}$ (dotted lines in Fig. 12c).
A conclusion from this comparison is that the rotational energy estimated by the flux-weighted {\NNDp} velocity gradient is much closer in value to the actual rotational energy of the core based on line of sight velocities.

Figure~\ref{fig:rottseries}d shows the time evolution of the gravitational energy, $E_{\rm grav}$, of the core. As with rotational energy, for observed cores one typically needs to make simplifying assumptions about the mass distribution of the core in order to estimate $E_{\rm grav}$. For the case of an uniform sphere, $E_{\rm grav}=(3/5) G M_c^2/R_c$. In the case of a singular polytropic sphere with density power law index $k_\rho$, $E_{\rm grav}= (3-k_\rho) G M_c^2/ (5-2k_\rho) R_c$ \citep{Bertoldi1992}. For the $k_\rho=1.5$ index that is relevant for the initial condition of our PSC, $E_{\rm grav}=(3/4)GM_c^2/R_c$, i.e., only a 25\% difference compared to the uniform sphere. This result is shown by the dotted lines in Figure~\ref{fig:rottseries}d for the two different mass estimation methods, i.e., with and without the density threshold along the line of sight (blue and red, respectively).

If the mass surface density map of a core is well resolved, we can use this distribution to carry out a pair by pair calculation for each pixel to estimate the gravitational energy, i.e., only using projected separations in the sky plane. Figure~\ref{fig:rottseries}d also shows this method, using dashed lines, for the simulated PSC. Since the distances between mass elements are underestimated in this method, $E_{\rm grav}$ is overestimated, i.e., compared to the singular polytropic sphere model estimate.

Finally, only in the case of a simulated PSC, we can use the full three-dimensional data to calculate the gravitational energy by pairing cells in 3D. This estimate is shown in Figure~\ref{fig:rottseries}d with the solid lines. We see that over most of the evolution of the simulation this method has close agreement with the polytropic sphere model estimate for the case where a density threshold is adopted (blue lines). However, at late times the 3D pairwise estimate becomes significantly higher, as the mass distribution deviates away from the simple polytropic model. For the method without the use a density threshold (red lines), we see that for most of the evolution both the polytropic model estimate and the 2D pairwise estimate are significantly larger than the true value yielded by the 3D pairwise method.

%The actual gravitational energy of the minimum core mass grows from about $8 \times 10^{43}$ to $3 \times 10^{45}$ erg.
%In contrast, the plane estimation grows from $1.3\times 10^{44}$ to about $5 \times 10^{45}$ erg (depending on the viewing direction). The polytropic model grows from $10^{44}$ to $10^{45}$ erg. It is shown that polytropic model gives good estimation in the first 0.5 {\tff}, and the error is still smaller than a factor of 3 even in the end. It is noticeable that no matter which way we used to estimate the potential energy, the energy given by minimum core mass will get closer to the one given by maximum core mass because they are dominated by the densest regions. 

% The huge gap in the beginning is caused by the limit of {\NNDp} emission because the area always lies in the center of the core and ignores the gas where $R > R_{c,eff}$. However, the outer layer do cause the core falls into a deeper potential and cause the gap. If we consider the whole core inside the $R_c$, they will have the same start point. Nevertheless, even if we ignore the effect, the discrepancy among the methods becomes even more apparent because the core is getting farther from the polytropic model and the spatial distribution is getting more complex.

In Figure~\ref{fig:rottseries}e, the ratio of rotational energy to gravitational energy is plotted. We focus on an estimate based on the {\NNDp}-weighted velocity gradient rotational energy using the 2D mass surface density map divided by the gravitational energy using the same 2D map. We find that the ratio $E_{\rm rot}/E_{\rm grav}$ starts at a level of $\gtrsim0.1$ when viewed along the $x$ and $y$ axes, but about an order of magnitude lower when viewed along the $z$ axis. This variation is due simply to stochastic sampling of large scale modes of the turbulent velocity field, since it is present at $t=0$. We note that the core was initialized with solenoidal turbulence, but with no large scale rotation beyond that arising from random sampling of this turbulent velocity field. In this sense the degree of rotation of these structures is the smallest it can be for such turbulent cores.

Then $E_{\rm rot}/E_{\rm grav}$ decays to a level of about 0.01 during the first 20 to 30 kyr, likely due to decay of the initial turbulent motions, but then shows an increase, especially when viewed along the $x$ axis, driven by infall motions. Finally there is a later decline to smaller values, driven in part by the late-time increase in $E_{\rm grav}$.

%We only plot the ratio of {\NNDp}-weighted rotational energy and the actual potential energy to represent the evolution.

% It is noticeable that the actual value decreases from 0.04 to around 0.008. If we use the weighted velocity gradient and the momentum of inertia estimated from the projected plane. Depending on the direction, they ranges from 0.002 to 0.03 in the end.

For polytropic models, the ratio of rotational energy and gravitational energy can be expressed as:
\begin{equation}
\frac{E_{\rm rot}}{E_{\rm grav}} = \frac{(3-k_\rho) M R^2 \omega^2 / [3 (5-k_\rho)]}{(3-k_\rho) GM^2/[(5-2k_\rho) R]} = \frac{(5-2k_\rho)\omega^2R^3}{3(5-k_\rho)GM}.
\label{eq:erot}
\end{equation}
We use this equation to estimate $E_{\rm rot}/E_{\rm grav}$ using unresolved core properties and the {\NNDp}-weighted velocity gradient, and also show this estimate in Figure~\ref{fig:rottseries}e. We find that this simple method of estimating the energy ratio follows that based on the 2D mass surface density map reasonably well (within a factor of about two), especially at earlier times (up to $\sim30\:$kyr).

% In this way, the values 0.008, 0.13, and 0.035 are given to the estimation.

% Assuming the cores are uniform sphere, equation~\ref{eq:erot} gives the ratio $\rm E_{rot}/E_{grav} = 0.0020, 0.0042, 0.0087$ on each plane at 0.8 $t_{ff}$. It can be noticed that the rotational energy got from x-y plane oscillating on a certain level, but the energy from other two plane transit from one to another plane. Considering the first moment map is influenced by the tracer, we repeat the same analysis by applying {\NNHp} to the same model (CR4x\_fDN100\_fDCO1000) and obtain total velocity gradients of 0.057, 0.055, 0.082 km/s/pc and $\rm E/U = 0.0017, 0.0038, 0.0089$ on each plane at 0.8 $\tff$. It has little effect to the results.

\subsubsection{Velocity Dispersion and Implied Virial State}
\label{sec:virial}

\begin{figure*}
    % \mbox{}\hfill
    % \begin{subfigure}[t]{0.45\linewidth}
    %     \includegraphics[width=\linewidth]{fig/velpanel_DN100.png}
    %     \subcaption{fDN100\_fDCO1000 model}
    %     \label{fig:veldistr_dn100}
    % \end{subfigure}
    % \hfill
    % \begin{subfigure}[t]{0.45\linewidth}
    %     \includegraphics[width=\linewidth]{fig/velpanel_CR4_DN100.png}
    %     \subcaption{CR4x\_fDN100\_fDCO1000 model}
    %     \label{fig:veldistr_cr4dn100}
    % \end{subfigure}
    % \hfill\mbox{}
    % \caption{(a) The velocity distribution weighted by {\NNHp} and {\NNDp} along x-axis and z-axis of fDN100\_fDCO1000 model. The top row is weighted by {\NNHp} and the bottom row is weighted by {\NNDp}. The left column is the distribution of $v_x$ and the right column is the distribution of $v_z$. The black line shows the original distribution of the snapshot at 0.8 $t_{ff}$. The transparent blue line shows the results considering thermal broadening at T = 15K. The green line includes both thermal broadening and hyperfine structure broadening effect of {\NNHp}(3-2) emission or {\NNDp}(3-2) emission. The bin width of the histograms is 0.1 km/s. (b) The velocity distribution weighted by {\NNHp} and {\NNDp} along x-axis and z-axis of CR4x\_fDN100\_fDCO1000 model. The arrangement of panels is the same as (a).}
    \centering
    \includegraphics[width=\linewidth]{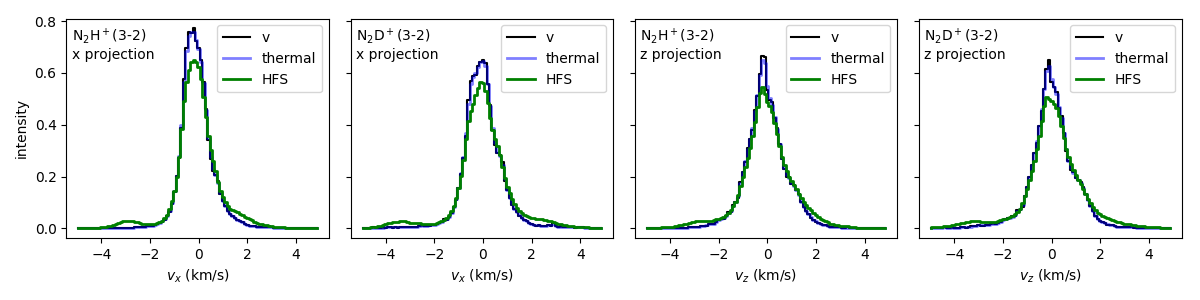}
    \caption{The velocity distribution weighted by {\NNHp} and {\NNDp} abundances along the x-axis and z-axis of fDN100\_fDCO1000 model. The left two panels show the distributions in $v_x$, weighted by {\NNHp} and {\NNDp} as labelled. The right two panels show the distributions in $v_z$ in the same format. The black lines show intrinsic velocity distributions of the snapshot at 0.8 $t_{ff}$. The transparent blue line shows the results including thermal broadening at $T = 15$~K. The green line includes both thermal broadening and hyperfine structure broadening of either {\NNHp}(3-2) emission or {\NNDp}(3-2) emission. The bin width of the histograms is 0.1 km/s.}
    \label{fig:veldistr}
\end{figure*}

In Figure~\ref{fig:veldistr}, we plot histograms of the velocity distribution of the CR4x\_fDN100\_fDCO1000 model viewed along the $x$ and $z$ axes at the final time step. The bin count in each cell is weighted by the abundances of either {\NNHp} or {\NNDp}. The histograms are normalized so that the area under each distribution equals unity. We plot three kinds of histogram. The first is the velocity distribution of the original data, i.e., the bulk motions of the gas due to turbulence, infall, rotation, etc. (black lines). The second is the total velocity distribution of bulk motions plus thermal broadening of the {\NNHp} or {\NNDp} species (transparent blue lines). For the assumed temperature of $T = $15~K, this thermal broadening is equivalent to a 1-D velocity dispersion of $\sigma \simeq 0.065\: {\rm km\:s}^{-1}$ for these species (note the 1-D sound speed in the gas given a mean particle mass of $2.33 m_{\rm H}$ is $0.230 (T/15\:{\rm K})^{1/2} {\rm km\:s}^{-1}$). The last histogram considers both the thermal broadening effect and the hyperfine structure of the {\NNHp}(3-2) emission or {\NNDp}(3-2) emission (green lines). 

All these distributions are dominated by a strong central peak. The effect of thermal broadening is very minor in comparison to the intrinsic bulk velocity dispersion. In contrast, hyperfine structure broadening acts to smooth and widen the distribution, but we note that a Gaussian fitted to the central part of the distribution still retains a similar width.

The velocity dispersions of the intrinsic distributions (black lines in Figure~\ref{fig:veldistr}) are $\sigma_{x,{\rm N2H+}}=0.685\:{\rm km\:s}^{-1}$, $\sigma_{x,{\rm N2D+}}=0.848\:{\rm km\:s}^{-1}$, $\sigma_{z,{\rm N2H+}}=0.859\:{\rm km\:s}^{-1}$ and $\sigma_{z,{\rm N2D+}}=0.983\:{\rm km\:s}^{-1}$. We see that the velocities along the $x$-axis have a narrower distribution and a higher peak. However, the velocity distribution along the z-axis does have a localized sharp central peak shape. Additionally, we see that {\NNHp} has a slightly narrower distribution than {\NNDp}. The reason for this is because we have assumed that {\NNHp} and {\NNDp} are emitted from the same volume (i.e., using the same density threshold). However, velocity dispersions are higher in the center due to infall motions, and this is where {\NNDp} is relatively enhanced with respect to {\NNHp}.
%is slightly reduced. Then the distribution weighted by the enhanced {\NNDp} shows a higher dispersion than the distribution weighted by the reduced {\NNHp}.} 

%{\bf some discussion about why there may be these small differences... i.e., N2D+ comes from denser regions undergoing faster collapse motions?}

% The evolution of the true velocity dispersions of the fDN100\_fDCO1000 and CR4x\_fDN100\_fDCO1000 models are shown in Fig.~\ref{fig:sigmatevolve}. In the $x$ and $y$ directions, the velocity dispersions drop down and then increase again after roughly 0.5 $t_{ff}$, but they always keeps increasing in the $z$ direction.

% In Figure~\ref{fig:virialtseries}a, we plot the time evolution of the velocity dispersion, as measured by the velocities along the three viewing directions. Here we measure velocity dispersion from the raw velocities of the {\NNDp}-emitting cells, weighted by their abundance of {\NNDp}. Note, to estimate this from an observed spectrum would require removing the contribution of thermal broadening and hyperfine structure broadening. Also, an observational study would need to correct for any optical depth effects in the observed spectrum.

\begin{figure*}
    \includegraphics[width=\linewidth]{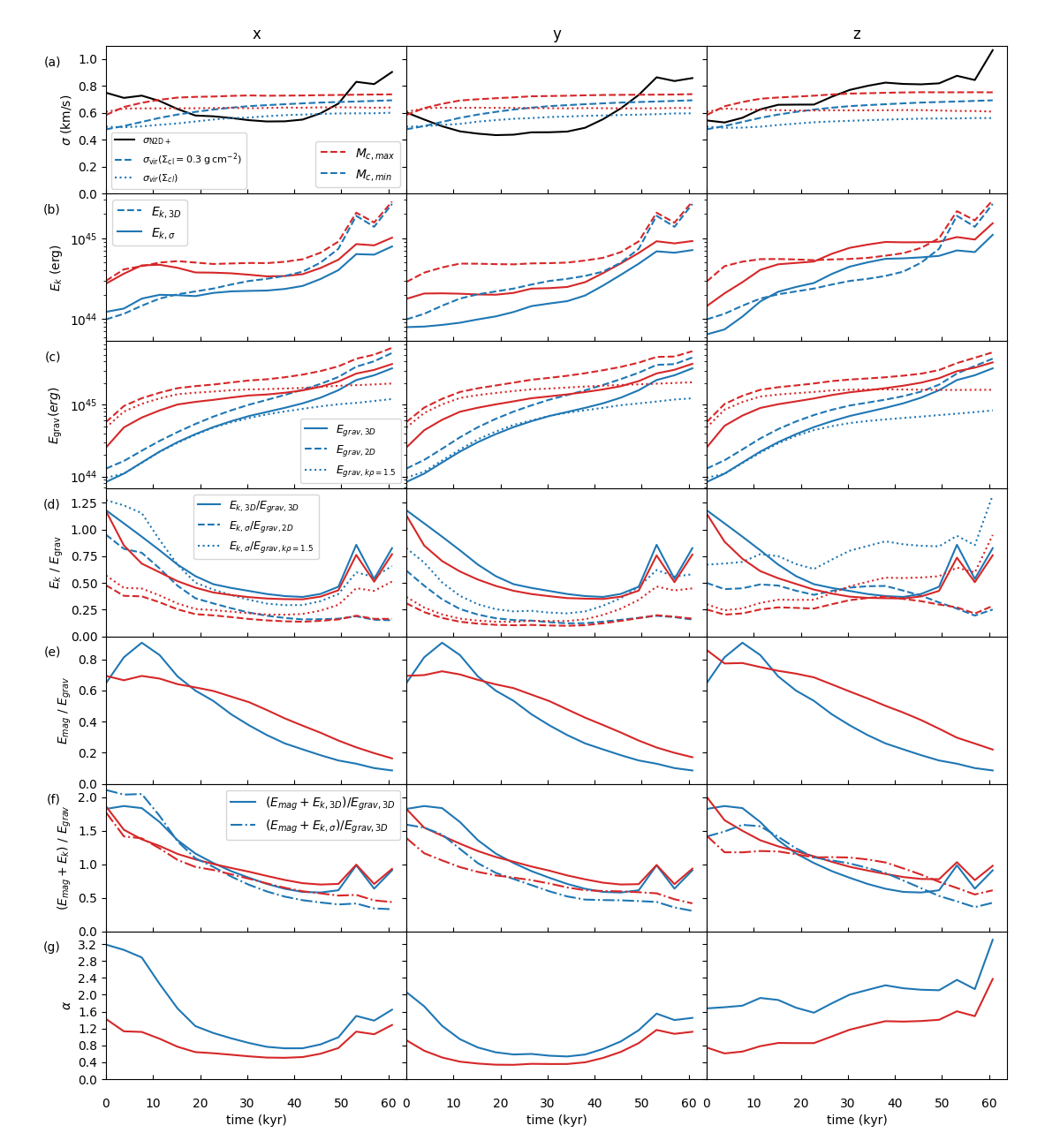}
    \caption{Time evolution of various dynamical properties of the PSC in the \mbox{CR4x\_fDN100\_fDCO1000} simulation, with results from the $x$, $y$, $z$ viewing directions shown in the columns left, middle, right, respectively. {\it (a) Top row:} Total one-dimensional velocity dispersion as traced by {\NNDp}, $\sigma_{\rm N2D+}$, including the thermal sound speed of $0.23\:{\rm km\:s^{-1}}$ added in quadrature (black solid lines). Also shown are estimates of the virial equilibrium velocity dispersion for two different assumptions about the clump mass surface density (a constant value of 0.3~$\rm g\:cm^{-2}$ - dashed lines; a local estimate around the core - dotted lines) and for two mass estimation methods of the core, i.e., with (blue) and without (red) density threshold along line of sight (see text). {\it (b) Second row:} Core kinetic + thermal energy, $E_k$, estimated from the 1D velocity dispersion traced by {\NNDp} (solid lines) and from the actual 3D motion in the simulation (dashed lines), again for the two mass cases described in (a) (blue and red).  {\it (c) Third row:} Core gravitational energy, $E_{\rm grav}$, estimated by (i) pairing cells in 3D data (solid lines), (ii) pairing cells in the 2D projected plane (dashed lines), (iii) a polytropic model with $k_\rho=1.5$ (dotted lines). {\it (d) Fourth row:} Ratio of kinetic + thermal energy to gravitational energy estimated by (i) kinetic + thermal energy and gravitational energy from 3D data (solid lines), (ii) kinetic + thermal energy estimated by $\sigma_{\rm N2D+}$ and gravitational energy estimated by pairing cells in the 2D projected plane (dashed lines), (iii) kinetic + thermal energy estimated by $\sigma_{\rm N2D+}$ and gravitational energy estimated by polytropic model with $k_\rho=1.5$ (dotted lines). {\it (e) Fifth row:} Ratio of core magnetic energy, $E_{\rm mag}$, to gravitational energy (see text). {\it (f) Sixth row:} Ratio of the sum of kinetic, thermal and magnetic energies to gravitational energy estimated by (i) kinetic + thermal energy and gravitational energy from 3D data (solid lines), (ii) kinetic + thermal energy estimated by $\sigma_{\rm N2D+}$ and gravitational energy from 3D data (dot-dashed lines). {\it (g) Bottom row:} Time evolution of virial parameter based on $\sigma_{\rm N2D+}$ 
    %the {\NNDp}-weighted velocity dispersion 
    and effective radius for the two mass estimation methods.}
    \label{fig:virialtseries}
\end{figure*}

In Figure~\ref{fig:virialtseries}a, we plot the time evolution of the total 1-D velocity dispersion, $\sigma_{\rm N2D+}$, as measured by the velocities of the {\NNDp}-emitting cells along the three viewing directions (black solid lines). Note that here, since we will use this velocity dispersion for a dynamical analysis, we include the thermal sound speed of $0.23\:{\rm km\:s}^{-1}$ added in quadrature to the bulk velocity dispersion as measured 
%, although it contributes much less than the turbulent motion. We first measure the velocity dispersion 
from the raw velocities of the {\NNDp}-emitting cells, weighted by their abundance of {\NNDp}.
%, and then add the contribution from sound speed at 15 K ( $\sim 0.23$ km/s).
Note, to estimate this from an observed spectrum would require removing the contribution of thermal and hyperfine structure broadening, before then adding in the 1-D sound speed. Also, an observational study would need to correct for any optical depth effects in the measured spectrum.

In the $x$ and $y$ directions, the velocity dispersion at first decreases modestly due to the decay of the initial turbulence. It then increases again after about 0.5 {\tff}. In the $z$ direction it exhibits a more monotonic increase. The velocity dispersion generally stays in a range of 0.4 to 0.8~km/s, except for a late increase seen in the $z$ direction, where it rises to just over 1.0~km/s.

In the context of the fiducial Turbulent Core model of \citet{2003ApJ...585..850M}, there is a predicted pre-stellar core velocity dispersion based on its mass and the mass surface density of the surrounding clump environment. The mass averaged velocity dispersion of the core in the fiducial case is given by \citep[see also][]{Tan2013}:
\begin{equation}
    \sigma_{\rm c,vir} = 1.09 \left(\frac{M_c}{60\:M_\odot}\right)^{1/4}\left(\frac{\Sigma_{\rm cl}}{1\:{\rm g\:cm}^{-2}}\right)^{1/4}\:{\rm km\:s}^{-1}.
    \label{eq:sigmavir}
\end{equation}
Note, this assumes the core is in virial and pressure equilibrium. As the core starts to undergo collapse and/or turbulence begins to decay, one expects deviations from the estimate of eq.~(\ref{eq:sigmavir}). 
If we adopt a fixed value of $\Sigma_{\rm cl}=0.3\:{\rm g\:cm}^{-2}$, which was used to set up the initial PSC, then we can use the evolving estimates of $M_c$ to calculate $\sigma_{\rm c,vir}$. 
% For this we use $M_{\rm c,max}$ (see Fig.~\ref{fig:rottseries}), which is closest to the method that can be done in observational studies. 
Virial velocity dispersion estimates using this method are shown in Figure~\ref{fig:virialtseries}a with dashed lines. Since we have two definitions of core mass, $M_{\rm c,min}$ and $M_{\rm c,max}$, the two estimates of velocity dispersion are plotted in blue and red, respectively. Alternatively, we can measure $\Sigma_{\rm cl}$ directly from the mass surface density map using the annular area of the region from $R_c$ to $2R_c$, which follows the method applied to analysis of observed cores by \citet{Tan2013} and \citet{Kong2017}. This estimate is also shown in Figure~\ref{fig:virialtseries}a (dotted lines). Again, we plot both estimates of $M_{\rm c,min}$ (blue) and $M_{\rm c,max}$ (red). We see that the actual value of $\sigma_{\rm N2D+}$ is quite similar to the estimates based on virial equilibrium. However, for much of the evolution and for two out of the three viewing directions, the core can appear kinematically sub-virial, i.e., with its observed velocity dispersion being lower than the virial equilibrium prediction by up to a factor of two. There is a tendency for velocity dispersion to rise above the virial equilibrium prediction in the final stages of the evolution of the PSC, i.e., just before the onset of star formation.

Observationally, the studies of \citet{Tan2013} and \citet{Kong2017} have identified 12 PSC candidates from their {\NNDp}(3-2) emission in IRDCs. The ratio of the observed {\NNDp}(3-2) velocity dispersion to that predicted by equation~(\ref{eq:sigmavir}) has been found to be equal to $0.83\pm0.15$ for the 6 cores studied by \citet{Tan2013} and $0.80\pm0.08$ for the 6 cores studied by \citet{Kong2017}. We see that such ratios are quite consistent with the results shown in Figure~\ref{fig:virialtseries}a. Thus, what is to be concluded here is that the overall velocity dispersion of {\NNDp} emission in the simulated PSC is always within a factor of two of the value predicted by virial equilibrium, even though the core is undergoing quite rapid collapse.

The total 1D velocity dispersion, $\sigma \equiv \sigma_{\rm N2D+}$, allows us to estimate the total internal kinetic + thermal energy of the core, assuming $\sigma_{\rm 3D}=3^{1/2} \sigma$. In Figure~\ref{fig:virialtseries}b, we plot the evolution of the kinetic + thermal energy estimated as $E_k = (3/2) M_c \sigma^2$ (solid lines), with the two different mass cases, i.e., with density threshold (blue) and without density threshold along the line of sight (red), as before. We also show the actual kinetic + thermal energy as estimated from the 3D velocities of the simulation data. The two methods generally track each other reasonably well, though with deviations that can be at the level of a factor of a few. We notice in the $y$ direction the velocity dispersion is relatively low during much of the evolution compared to other directions, so this leads to a relatively larger deviation given the assumption of isotropic velocity dispersion that is made in the simple estimate. At late times, the actual kinetic + thermal energy in the core increases significantly, which we attribute to the emergence of a modest number of very dense cells, with deep gravitational potentials that induce fast motions.

%For the minimum core mass, the theoretical value shows monotonic increase from several $10^{43}$ to about $10^{45}$ erg (depending on the viewing direction). The maximum core mass gives another value which is always a little bit higher because of the mass ratio. In contrast, the simulation data shows the actual kinetic energy grows from about $9 \times 10^{43}$ to about $3 \times 10^{45}$ erg in the case of minimum core mass. The maximum core mass has a larger value in the beginning but converges to similar value in the end. The values given by the two ways does not show an apparent difference in the beginning and shows different variation depending on the viewing direction, but the actual kinetic energy is always larger than the theoretical value in the end.

In Figure~\ref{fig:virialtseries}c, we plot the gravitational energy of the core material that is traced by {\NNDp}, which is a repeat of the information shown in Figure~\ref{fig:rottseries}d, shown here for convenience. Then in Figure~\ref{fig:virialtseries}d we plot the evolution of the ratio of kinetic + thermal to gravitational energies, i.e., $E_k/E_{\rm grav}$.
%The ways to estimate the gravitational energy have been described in the last subsection. 
The first ratio to consider is that of the actual kinetic + thermal energy estimated from 3D motions to that of the gravitational energy estimated from a pairwise cell calculation in 3D (solid lines) (see \S\ref{sec:firstmom}). This ratio starts with relatively high values, which can be $>1$, and then declines, which is due to the decay of the initial, supervirial turbulence.
% The ratio of these energies {\color{red} stabilises} at a level of about 0.2 at late times, which is smaller than the value of 0.5 expected from a simple application of the virial theorem for a non-magnetised core without consideration of surface pressure terms (see below).
The ratio declines to about 0.4, which is close to virial equilibrium, and then bounces back to a range between 0.5 to 0.8. 

Other methods of estimating $E_k/E_{\rm grav}$ are those that can be done observationally, i.e., with $E_k$ estimated from $\sigma$ of {\NNDp} and $E_{\rm grav}$ from either a 2D mass surface density map (i.e., 2D pairwise summation) (dashed lines) or from the simple spherical polytropic assumption (dotted lines). 
If we consider the density threshold (blue lines), the 2D method yields an estimate generally within a factor of two of the actual 3D result, but the error could be larger in the later stages.
% except in the $y$ projection, where, as we have noted, the velocity dispersion tends to be relatively low leading a more severe underestimate of $E_k/E_{\rm grav}$. 
The polytropic estimate yields a slightly better accuracy, within about a factor of two of the 3D result. When the density threshold is not used, then the mass is overestimated, leading to lower values being inferred for $E_k/E_{\rm grav}$.

The magnetic energy of the core is expected to play an important dynamical role in the evolution of the core. We calculate $E_{\rm mag} = \int B^2 / (8\pi) dV$, with the integration approximated as a summation over the volume of the cells that make up the volume of the core. The ratio $E_{\rm mag}/E_{\rm grav}$ is shown in Figure~\ref{fig:virialtseries}e. For the case with the core mass defined with the density threshold (blue line) there is a short rising phase in the first 10~kyr when this ratio approaches unity. After this there is a decline to values approaching $\sim0.1$. For the core defined without the density threshold used along the line of sight (red lines), the ratio shows a simpler, monotonic decline from about 0.7 to about 0.2.

The ratio of total internal energies ($E_k+E_{\rm mag}$) to gravitational energy is shown in Figure~\ref{fig:virialtseries}f. For the actual energy ratio, it shows a declining evolution from values $\sim2$ down to a range of 0.5 to 1. Applying the kinetic + thermal energy estimated from $\sigma_{\rm N2D+}$ gives an even monotonic decline down to a range of 0.3 to 0.5.

% As expected, this shows a declining evolution from values $>1$ down to values that are $\sim0.3$ to 0.5, {\color{red} with a quite slow change} during the later stages of the simulation, i.e., after $\sim30\:$kyr.

%mainly use the gravitational energy estimated from pairing cells in three-dimensional cube as a standard to calculate the ratio of kinetic energy and gravitational energy. Considering the kinetic energy given by theoretical model, the ratio shows a decreasing trend. The initial value ranges from 0.4 to 1.3 (depending on the viewing direction) and decreases to about from 0.2 to 0.2 in the end. In contrast, the actual ratio decreases from 1.0 to about 0.4 in the first 0.6 {\tff} and then grows up to 0.8 in the end.

In Figure~\ref{fig:virialtseries}g, we plot the virial parameter, $\alpha \equiv 5 \sigma^2 R_c/ (G M_c)$, which is often used in observational studies and is equivalent to twice the ratio of kinetic energy and gravitational energy in the uniform sphere case.
%The variation of the virial parameter is similar to the velocity dispersion. 
The virial parameter decreases in the first 0.5 {\tff} in the $x$ and $y$ directions, but keeps increasing in the $z$ direction. Its value ranges from 0.3 to 3.0. For much of the evolution the virial parameter can take values that are $<1$, so that a core would be interpreted as being kinematically "sub-virial". We see that the value also depends on how the mass is measured in the core, i.e., illustrating the importance of background subtraction methods to remove the contribution of material along the line of sight.

\begin{figure*}
    \centering
    \includegraphics[width=\linewidth]{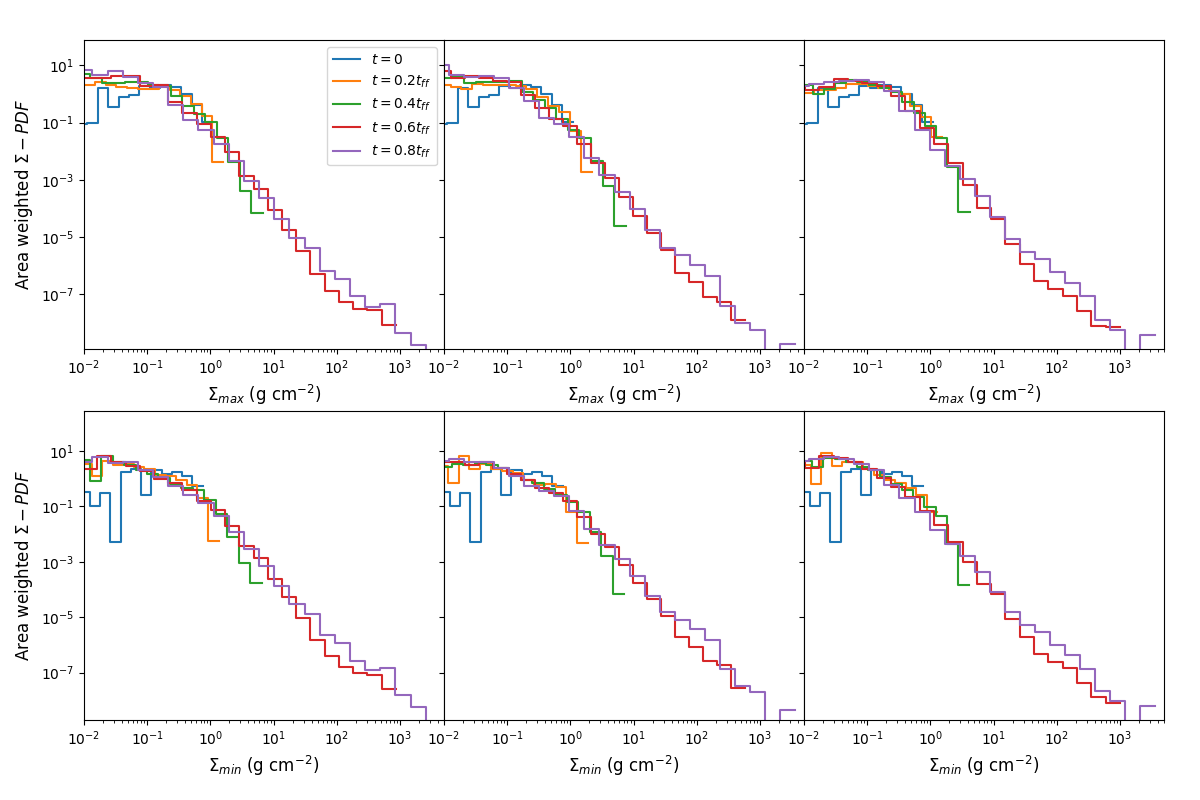}
    \caption{Probability distribution functions of mass surface density. {\it (a) Top row:} Distribution of mass surface densities including cells along the line of sight of the emission area. The results for $x$, $y$ and $z$ projections are shown in the left, middle and right panels, respectively. The different colors show the time evolution of the distributions (see legend). {\it (b) Bottom row:} As (a), but now for mass surface density only of the cells where $\nH > \nHcrit$. }
    \label{fig:masspdf}
\end{figure*}

Finally, as another metric to examine the collapsing state of the core, we plot the mass surface density distribution function (area weighted) in Figure~\ref{fig:masspdf}. A collapsing core is expected to develop regions of very high values of mass surface density, which could serve as a diagnostic of the evolutionary state of the PSC. From the distributions shown in Figure~\ref{fig:masspdf}, we do see the steady development of a high-$\Sigma$ tail to the distribution, however the shape is relatively constant in the range from 0.1 to 100~$\rm g\:cm^{-2}$ from $t=0.6$ to $0.8\:t_{\rm ff}$. 

Observationally, mid-infrared extinction mapping methods \citep[e.g.][]{Butler2012, Butler2014} based on {\it Spitzer}-IRAC 8~$\rm \mu m$ imaging data of Galactic plane infrared dark clouds have relatively poor angular resolution of 2\arcsec and tend to saturate at values of $\Sigma\sim 0.5$ to 1~$\rm g\:cm^{-2}$. Thus such maps are not currently able to probe most of the dynamic range shown in the $\Sigma$-PDFs of Figure~\ref{fig:masspdf}. High angular resolution observations of sub-mm/mm dust continuum emission, e.g., with {\it ALMA}, are another method to probe the $\Sigma$-PDFs of PSCs, but one would also need to allow for the potential effects of temperature and (grain growth/composition driven) opacity variations \citep[e.g.,][]{Lim2015, 2017A&A...600A.123D, 2017A&A...606A..50D} in the core.

%{\bf virial analysis, following method of Tan et al. (2013)... basically compare expected velocity dispersion from Turbulent Core Model to that seen in the simulations, as a function of time, and viewed in different directions. Make note of our initial conditions, which probably had extra turbulence compared to virialized state... evolution is probably due to decay of initial turbulence, followed by increase in velocities due to collapse.}

% \begin{figure}
%     \includegraphics[scale=0.5]{fig/CR10_velx.png}
%     \caption{The distribution of velocity weighted by {\NNHp} and {\NNDp}. The projection of velocity is along x-axis. The projection is considered in a $0.2\times 0.2 (pc^2)$ area (the same as the area plotted in Fig.~\ref{fiducial_x} and ~\ref{fiducial_z}). The bin width of the histograms is 0.1 km/s. }
% \label{CR10_velx_w1d2}
% \end{figure}

% \begin{figure}
%     \includegraphics[scale=0.5]{fig/sigmapanel.png}
%     \caption{The one-dimensional velocity dispersion weighted by {\NNHp} and {\NNDp} along three axes. The top panel shows the results from CR10x model and the bottom panel shows the results from fDCO1000 model.}
%     \label{fig:sigmatevolve}
% \end{figure}

\section{Summary and Conclusions}
\label{sec:discussion}
% \subsection{Chemical model}

We have presented a simulation of a massive, magnetized PSC, i.e., the starting assumption of the Turbulent Core Model of massive star formation \citep{McKee2002, 2003ApJ...585..850M}. The main novel aspect is the self-consistent coupling of a detailed astrochemical gas phase network that follows the evolution of a number of species that can trace dense, cold gas, especially {\NNHp} and {\NNDp}. The latter species in particular has been used as a diagnostic tracer of massive PSCs \citep[e.g.,][]{Tan2013, Kong2017, Kong2018}. This simulation work extends the methodology of a previous study of \citet{Goodson2016}, which only approximated the chemistry with semi-analytic growth functions for the abundances of these species and which was a lower resolution, uniform grid simulation. One of the main differences of this work compared to that of \citet{Goodson2016} is that we predict lower levels of deuteration. The difference is because the previous semi-analytic method does not explicitly follow the time evolution of other species, especially ortho- and para- {\HH}. Furthermore, this previous method was not able to fully follow advective diffusion in a turbulent core.
% cjhsu - The main reason should be losing information of other species.
% not being able to fully follow advective diffusion in a turbulent core.

With the methodology developed here, we first investigated a fiducial astrochemical model and found that only relatively low levels of deuteration could be achieved during the approximately free-fall collapse, i.e., only reaching {\DfracNNH}$\sim0.002$, which is much lower than found in many observed PSCs. We then carried out an extensive survey of different astrochemical model parameters and properties, i.e., initial ortho-to-para ratio of $\rm H_2$, cosmic-ray ionization rate, heavy element depletion factors, degree of chemical pre-aging, to investigate how high levels of deuteration, approaching unity, can be achieved, which are needed to explain the observed PSCs. We have found that a model with relatively high cosmic-ray ionization rates ($1.0\times10^{-16}\:{\rm s}^{-1}$) and high depletion factors of CO (1000) and N (100) is able to match observed properties of some PSCs, i.e., {\DfracNNH} $\sim0.1$ to 1 and absolute abundances of these species, within the followed evolution of the core, i.e., within $0.8\:t_{\rm ff} = 61$~kyr. However, it should be emphasised that this is just an example model and other combinations of parameters are expected to also be able to give similar agreement with these observational metrics. One additional metric that may help break degeneracies is the spatial structure of {\DfracNNH}, i.e., how it correlates with column density. However, to date such observations have not yet been acquired to be able to test the models.

In kinematic and dynamic aspects, we examined the capability of {\NNDp} as a tracer. We used the {\NNDp} abundances to imitate observational analyses. This showed that the velocity gradients estimated by {\NNDp} give a good estimation of the actual rotational to gravitational energy ratio based on the assumption of a polytropic core structure model. We also found that the PSC, as defined by its {\NNDp} emission, can often appear kinematically sub-virial, even though it is undergoing quite rapid collapse. High resolution observations of the velocity structures of massive PSCs as traced by {\NNDp} are one promising method to help test models of PSC evolution. From the simulation side, a broader range of physical models, e.g., with a varying degree of $B$-field support and with a range of initial conditions that set core formation, are needed to explore their effects on these kinematic signatures.

%\subsection{Time dependent {\CO} depletion model}

%\subsection{Caveats of our analysis}

Some caveats of our work include that, for simplicity, we do not include freeze-out and desorption reactions and surface chemistry. Instead we control the gas phase C, O and N abundances by parameterized depletion factors. However, the depletion rate of {\CO} is dependent on density. The {\CO} depletion factor is likely to have a spatial structure. In our fDCO100 and fDCO1000 models, it is obvious that they reach different levels of deuterium fractionation. If there is a spatial structure of {\CO}, this could influence the final result of our simulation. In \citet{Bovino2019}, they discuss the {\CO} depletion structure of their simulated high-mass pre-stellar region. Their result shows that the {\CO} depletion does have a spatial structure and thus influences the structure of deuterium fractionation also. Thus an additional observation metric of {\CO} depletion structure is likely to be helpful to give better constraints on our PSC models, which ultimately should include time dependent depletion and desorption processes and surface chemistry, although these aspects introduce many additional uncertainties for the modeling. Another, similar caveat is our choice of uniform cosmic-ray ionization rate, rather than allowing for attenuation of this rate in higher density, more shielded regions. Such effects will also be investigated in future work that extends the analysis of this study.

\section*{Acknowledgements}
% The Acknowledgements section is not numbered. Here you can thank helpful
% colleagues, acknowledge funding agencies, telescopes and facilities used etc.
The simulations were performed on resources provided by the Swedish National Infrastructure for Computing (SNIC) at C3SE. JCT acknowledges support from VR grant 2017-04522 (Eld ur is) and ERC Advanced Grant 788829 (MSTAR).

\section*{Data Availability}
The data underlying this article will be shared on reasonable request to the corresponding author.

%%%%%%%%%%%%%%%%%%%%%%%%%%%%%%%%%%%%%%%%%%%%%%%%%%



\bibliographystyle{mnras}
\bibliography{example} % if your bibtex file is called example.bib

% \begin{thebibliography}{99}
% \bibitem[\protect\citeauthoryear{Author}{2012}]{Author2012}
% Author A.~N., 2013, Journal of Improbable Astronomy, 1, 1
% \bibitem[\protect\citeauthoryear{Others}{2013}]{Others2013}
% Others S., 2012, Journal of Interesting Stuff, 17, 198
% \end{thebibliography}

%%%%%%%%%%%%%%%%%%%%%%%%%%%%%%%%%%%%%%%%%%%%%%%%%%

% %%%%%%%%%%%%%%%%% APPENDICES %%%%%%%%%%%%%%%%%%%%%

\appendix
\section{Comparison of chemodynamical methods}
\label{overestimateissue}

\begin{figure}
\centering
\includegraphics[scale=0.5]{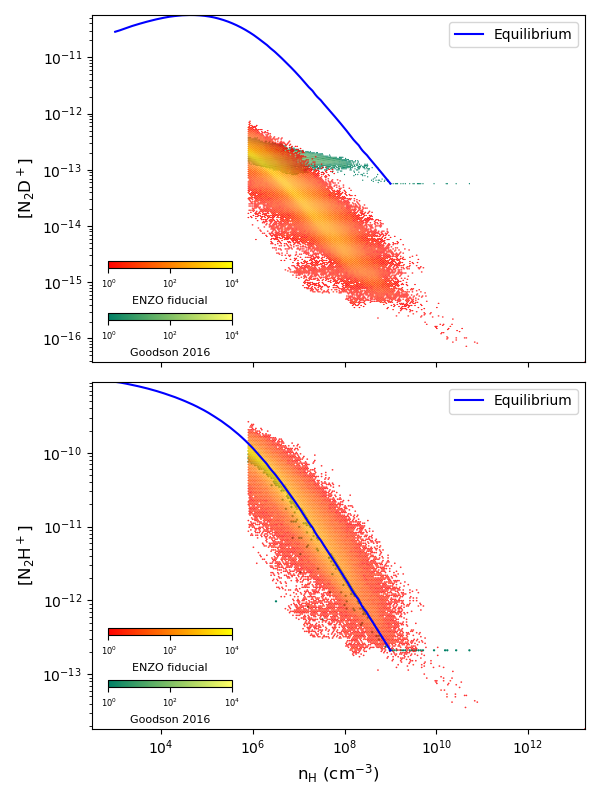}
% \subfigure{
%     \includegraphics[scale=0.5]{fig/scatter3d_N2D_t8.png}
% }
% \subfigure{
%     \includegraphics[scale=0.5]{fig/scatter3d_N2H_t8.png}
% }
\caption{The distribution of {\NNDp} and {\NNHp} versus column density. The blue line indicates the equilibrium value used in \citet{Goodson2016}. {\NNHp} (lower panel) matches with this equilibrium value well in both simulations, but the behaviour of {\NNDp} is very different between the two simulations: i.e., in our models gas at high density has a much lower abundance of {\NNDp}.}
\label{n2dn2hdistribution}
\end{figure}

\begin{figure}
\centering
\includegraphics[scale=0.5]{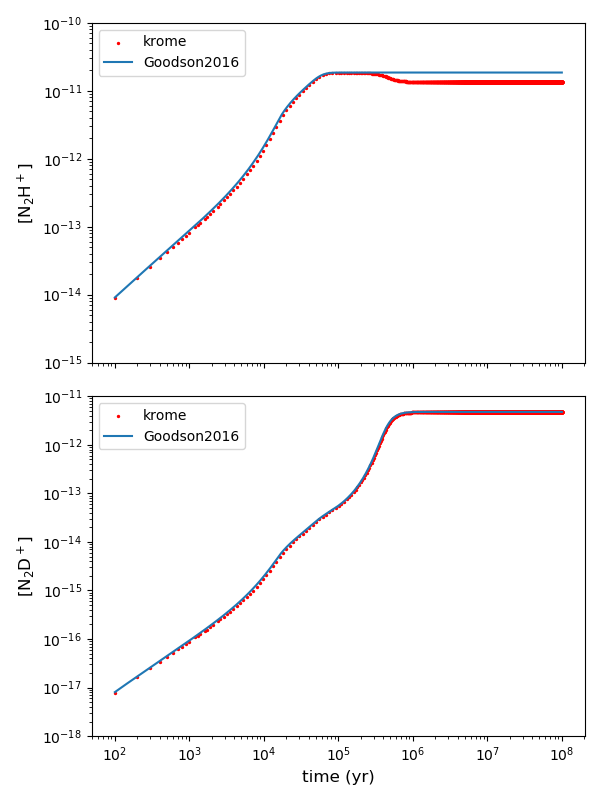}
% \subfigure{
%     \includegraphics[scale=0.4]{fig/n2d_d1e7.png}
% }
% \subfigure{
%     \includegraphics[scale=0.4]{fig/n2d_d1e9.png}
% }
% \subfigure{
%     \includegraphics[scale=0.4]{fig/n2h_d1e7.png}
% }
% \subfigure{
%     \includegraphics[scale=0.4]{fig/n2h_d1e9.png}
% }
\caption{We run single grid models and compare the results with the interpolation data of \citet{Goodson2016}, which corresponds to our fiducial model. The parameters have been listed in the Table~\ref{tab:chemmodel}. In single cells, they have a good agreement. We notice that the time scale required for $\NNHp$ to achieve the equilibrium value is shorter and is similar to one free-fall time. By contrast, the timescale for $\NNDp$ to achieve equilibrium value is longer than one free-fall time.}
\label{singlegridcheck}
\end{figure}

Comparing the results of our fiducial case with the corresponding results in \citet{Goodson2016}, we notice an apparent discrepancy in the estimated deuteration level. The peak value of deuterium fraction is about 100 times smaller in our simulation. We check the abundances of $\NNHp$ and $\NNDp$ and notice that $\NNHp$ has a good agreement but $\NNDp$ was much larger in \citet{Goodson2016} compared to our results (Figure~\ref{n2dn2hdistribution}). The agreement of {\NNHp} comes from the equilibrium condition. Since the timescale for {\NNHp} to achieve an equilibrium abundance is shorter than one free-fall time, both simulations reach near equilibrium values and match with each other very well. In contrast, the timescale for {\NNDp} to reach its equilibrium state is much longer. It is impossible for {\NNDp} to reach the equilibrium value in one free-fall time (Figure~\ref{singlegridcheck}). However, we notice {\NNDp} reaches the equilibrium value in the densest grids in \citet{Goodson2016} (Figure~\ref{n2dn2hdistribution}). The reason why {\NNDp} reaches the equilibrium state in a much shorter time is a result of the approximate interpolation method. The growth rate of {\NNDp} and {\NNHp} are estimated by their current abundances respectively. The interpolation method ignores the information of other species, such as {\OPRHH}, which suppress the growth of {\DfracNNH}.

\begin{figure}
\centering
    \includegraphics[scale=0.5]{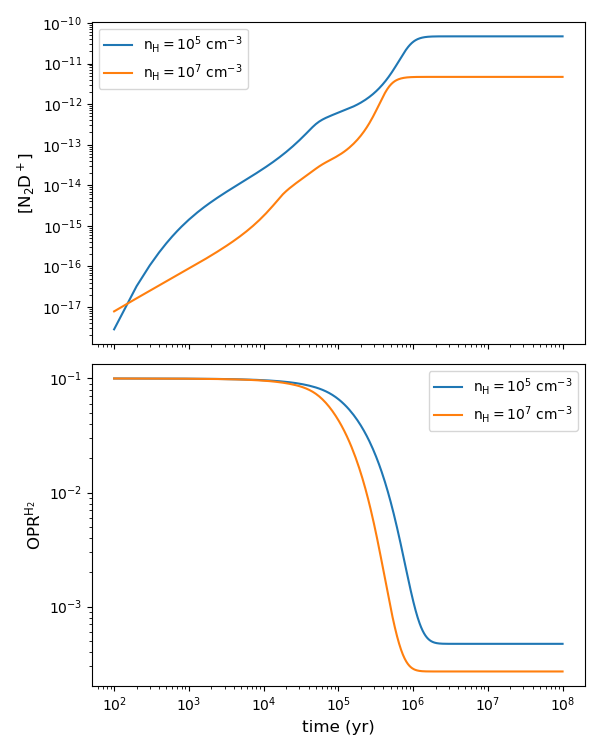}
\caption{The evolution of {\NNDp} in single grid models. We use two different densities here: $10^5$ and $10^7\ \mathrm{cm}^{-3}$.}
\label{n2d_opr_evolve}
\end{figure}

Considering two single grid models, which have the same initial parameters as the fiducial model (see Table~\ref{tab:chemmodel}), with $n_{\rm H} = 10^5$ and $10^7\:{\rm cm}^{-3}$, their {\NNDp} growth curves are plotted in Figure~\ref{n2d_opr_evolve}. If $\NNDp$ is $10^{-13}$ in one grid, the grid could have evolved for around either 27,000 years in the former case or 160,000 years in the latter case. It means they have different history and different chemical ages. \citet{Goodson2016} used the local abundance to predict the growth of {\NNHp} and {\NNDp}. 
%This cause the ignorance of the history. 
%jctfinal - there seem to be mistakes in the numbers here... how can OPR be 0.9????????
If we check {\OPRHH} in Figure~\ref{n2d_opr_evolve}, we even find that they corresponds to $\OPRHH \approx 0.091$ and $0.023$ respectively. Goodson et al. predicted the growth of {\NNDp} according to the point data where $\OPRHH = 0.023$, but the real {\OPRHH} is around 0.091 and this suppresses the growth of {\NNDp}. Since the high-density region (e.g., $n_{\rm H} = 10^7\:{\rm cm}^{-3}$) inherits the value from the low-density (e.g., $n_{\rm H} = 10^5\:{\rm cm}^{-3}$) region, it finally gives an overestimated result.
% \begin{figure}
% \centering
%     \includegraphics[scale=0.4]{fig/n2dgrowth.png}
% \caption{The evolution of {\NNDp} in single grid models. We use two different densities here: $10^5$ and $10^7\ \mathrm{cm}^{-3}$ }
% \label{n2dgrowth}
% \end{figure}
% \begin{figure}
% \centering
%     \includegraphics[scale=0.4]{fig/oprh2evolve.png}
% \caption{The corresponding evolution of {\OPRHH} in the models showed in Fig.~\ref{n2dgrowth}.}
% \label{oprh2evolve}
% \end{figure}

\section{Elemental conservation}
\begin{figure}
    \centering
    \includegraphics[width=\linewidth]{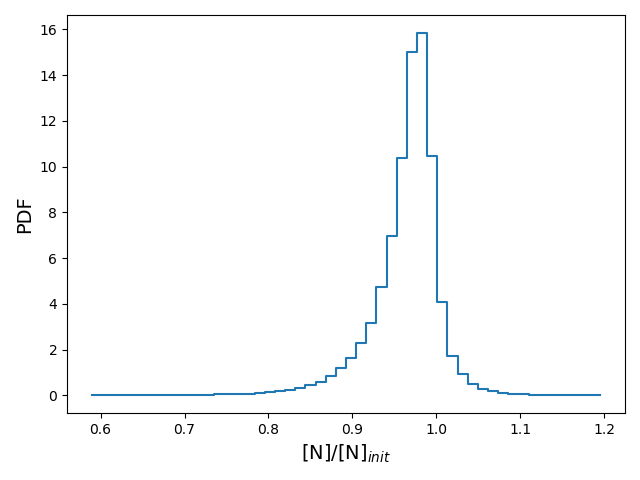}
    \caption{The ratio of [N]/[N]$_{\rm init}$ of cells in the defined core.}
    \label{fig:histN}
\end{figure}

In adaptive mesh refinement (AMR) frameworks, interpolation can cause errors in elemental conservation because of gradient limiters and interpolation weights could be different for each species \citep{Plewa1999, Grassi2017}. This means that the total abundances of C, N and O would not be conserved, even if each individual chemical species is conserved during the interpolation. To check elemental conservation, we compare the total abundances of C, N and O with their initial values cell by cell. We find that their mean values have a typical $\sim$2\% error for each element in our defined cores. For example, Figure~\ref{fig:histN} shows the ratio of [N]/[N]$_{\rm init}$. The mean value is 0.977 and the standard deviation is 0.021. Other elements, i.e., C and O, have similar distributions. The error is even smaller if we consider a more extensive domain because the errors are concentrated in the refined regions within our defined cores. Such levels of elemental conservation errors do not significantly influence our results. 
%but {\KROME} package provides a solution to keep the conservation law \citep{Grassi2017}. 

% \appendix

% \section{Some extra material}

% If you want to present additional material which would interrupt the flow of the main paper,
% it can be placed in an Appendix which appears after the list of references.

% %%%%%%%%%%%%%%%%%%%%%%%%%%%%%%%%%%%%%%%%%%%%%%%%%%

% Don't change these lines
% \bsp	% typesetting comment
\label{lastpage}
\end{document}